# Implementing Reactivity in Molecular Dynamics Simulations with Harmonic Force Fields

by

Jordan J. Winetrout,[1,2] Krishan Kanhaiya,[1,2,3] Joshua Kemppainen,[4] Pieter J. in 't Veld,[5] Geeta Sachdeva,[6] Ravindra Pandey,[6] Behzad Damirchi,[7] Adri van Duin,[7] Gregory M. Odegard,[4] Hendrik Heinz[1,2*]

[1] Department of Chemical and Biological Engineering, University of Colorado at Boulder, Boulder, CO 80309, USA

[2] Materials Science and Engineering Program, University of Colorado at Boulder, Boulder, CO 80309, USA

[3] Insitute of Physics, Ruhr University Bochum, Universitätstrasse 150, 44801 Bochum, Germany

[4] Department of Mechanical Engineering - Engineering Mechanics, Michigan Technological University, Houghton, MI 49931, USA

[5] BASF SE, Molecular Modeling & Drug Discovery, 67056 Ludwigshafen, Germany

[6] Department of Physics, Michigan Technological University, Houghton, MI 49931, USA

[7] Department of Mechanical Engineering, Pennsylvania State University, University Park, PA 16802, USA

* Corresponding author: hendrik.heinz@colorado.edu

**Abstract**

The simulation of chemical reactions and mechanical properties including failure from atoms to the micrometer scale remains a longstanding challenge in chemistry and materials science. Bottlenecks include computational feasibility, reliability, and cost. We introduce a method for reactive molecular dynamics simulations using a clean replacement of non-reactive classical harmonic bond potentials with reactive, energy-conserving Morse potentials, called the Reactive INTERFACE Force Field (IFF-R). IFF-R is compatible with force fields for organic and inorganic compounds such as IFF, CHARMM, PCFF, OPLS-AA, and AMBER. Bond dissociation is enabled by three interpretable Morse parameters per bond type and zero energy upon disconnect. Use cases for bond breaking in molecules, failure of polymers, carbon nanostructures, proteins, composite materials, and metals are shown. The simulation of bond forming reactions was included via template-based methods. IFF-R maintains the accuracy of the corresponding non-reactive force fields and is about 30 times faster than prior reactive simulation methods.

## Introduction

The prediction of chemical reactions and mechanisms for bond breaking, failure, and toughness of advanced materials from the atomistic scale to the microstructural and macroscopic length scale is a grand challenge in materials science.[1-4] The interplay of chemistry, defects, cross-linking in polymers, hierarchical assembly from the molecular scale to the macroscale and multi-scale dynamics of materials generates a wide range of chemical, optical, conductive, thermal, elastic, plastic, and failure properties. The development of bioinspired and engineered functional materials is limited by difficulties in probing associated patterns of chemical reactions and stress response mechanisms using experimental and computational methods.[4-8] Simulation methods, such as molecular dynamics (MD) simulations, can accelerate understanding of the underlying mechanisms from the scale of atoms to micrometers, provide guidance in materials selection and processing. For example, the role of compositions, defects, reactions, and processing conditions has been explored utilizing feedback loops from experiment to simulation and back to experiment to shorten discovery and development times.[9-18] Current limitations include the modeling chemical reactions, restrictions in size (<μm) and computation time (<ms). In this contribution, we introduce a method for modeling chemical reactions, especially bond breaking and failure of materials, without sacrificing the high accuracy, relatively short computation time and large system size in classical MD simulations.

Many current force fields for MD simulations in the materials and biomolecular community use harmonic bond energy terms, in addition to other bonded and nonbonded terms, and are limited in predictions of bond-breaking and bond-forming reactions (IFF,[5] PCFF,[19] AMBER,[20] CVFF,[21] CHARMM,[22] OPLS-AA[23], COMPASS[24], DREIDING[25]).[14,26] Elements and specific chemistries are represented by atom types and often include multiple atom types for the same element. In some instances, such as metals, oxides, and ionic solids, only nonbonded terms are necessary and reactivity can be implemented with relative ease.[15,17,27] For example, alloying,[10,28] cis-trans photoisomerization reactions,[14] thiol-metal surface linking,[16,29] and hydration reactions[18,30-32] can be sufficiently described without the need for explicit changes in bonding. The dissociation and formation of covalent bonds, however, requires new approaches. To introduce reactivity for this class of potentials, we consider the INTERFACE force field (IFF) which, among all-atom bonded force fields, covers the most diverse chemical space across the periodic table in highest accuracy and compatibility.[5] IFF has been developed to exclusively use interpretable parameters, accurately represent chemical bonding, and reproduce the structural as well as the energetic properties of included compounds under standard conditions relative to experimental data and theory, exceeding DFT methods up to an order of magnitude in accuracy.[17,33,34] IFF was conceived to be transferable and compatible with existing harmonic force fields for organic molecules and biopolymers by using the same potential energy terms, including 12-6 and 9-6 options for the Lennard-Jones potential.[5] IFF covers, for example, metals, minerals, oxides, 2D materials, common water models, solvents, gases, organic compounds, and performs exceptionally well for mixtures, solid-electrolyte interfaces, and hybrid materials. Lattice parameters and densities of pure phases usually deviate <0.5%, surface and hydration energies <5%, and mechanical properties <10% from reference data, which is about one to two orders of magnitude more accurate than other force fields for the same compounds.[5,17,27,34] The reliability exceeds common electron density functionals[35-39] especially with regard to simulating interfaces.[17,34] IFF can be merged with biomolecular force fields such as CHARMM[22], AMBER[20], OPLS-AA[23], as well as PCFF[19,40] and COMPASS[24], extending access to protein, DNA, lipid, and carbohydrate based multifunctional materials.

To-date, bond order potentials such as the Reactive Empirical Bond Order (REBO) Potential (REBO)[41] and the Reactive Force Field (ReaxFF)[42] have been developed to address bond dissociation in large-scale MD simulations, requiring a high number of additional energy terms. Specifically, ReaxFF uses Pauling's bond order concept to describe the strength of a bond based on its chemical environment and interatomic distances.[42] ReaxFF can predict bond breakage and accurately reproduce reaction barriers and reaction energies derived from quantum mechanics for certain reactions.[43] However, the potential function contains more than triple the amount of energy terms compared to IFF or CHARMM, including extensive bond order terms, over-coordination, angle, dihedral, conjugation, lone pair, H-bond effects, and correction terms.[44] These terms add the ability to describe complex, concerted reaction paths and, at the same time, come at the cost of significant complexity, which can be difficult to interpret and which increase the computational expense. In contrast to IFF and biomolecular force fields, ReaxFF employs only one atom type for each element. This choice greatly simplifies the simulation setup by not requiring atom typing, however, it is also challenging for the force field developer and users. For example, the simulation of multiphase materials such as oxides in contact with water and oxygen-containing polymers requires multiple oxygen types. Therefore, different ReaxFF 'branches' have been developed, such as the combustion and the aqueous phase branches.[43] ReaxFF also does not cover complex polymers such as proteins, DNA, and their interfaces with various engineering materials with relatively high accuracy,[45] and it is desirable to combine the simplicity of a harmonic potential with the capabilities of ReaxFF.

As chemical reactions are highly sensitive to bond polarity, molecular geometry, and interaction energies, IFF is uniquely prepared to incorporate reactive processes. In this contribution, we augment IFF into the reactive Interface force field, IFF-R, to facilitate simulations of bond breaking and formation reactions. IFF-R uses a Morse potential to quantitatively represent the bond energy between pairs of atoms $i$ and $j$ as a function of distance, as opposed to a harmonic potential in IFF. The Morse representation aligns with experimentally measured energy functions and has a quantum mechanical justification. (Figure 1a).[46,47] Accordingly, we substituted the harmonic bond energy term in the energy expression with a Morse bond energy term, including either all types of bonds or a subset thereof, e.g., only bonds of lowest energy known to dissociate before others (Figure 1b, c). We demonstrate that this clean replacement requires no changes in other force field parameters to add bond breaking capabilities, maintains the full benefits of the non-reactive IFF, and speeds up reactive simulation ~30 times relative to existing methods. We also demonstrate the formation of bonds, or reversibility of bond breaking with IFF-R using template-based reaction simulations, specifically, the REACTER toolkit.[48] The formation and breaking of bonds is thus possible in a continuous simulation without the need for the more complex bond-order potentials. We demonstrate the application of IFF-R with PCFF and CHARMM functional forms (Figure 1d).

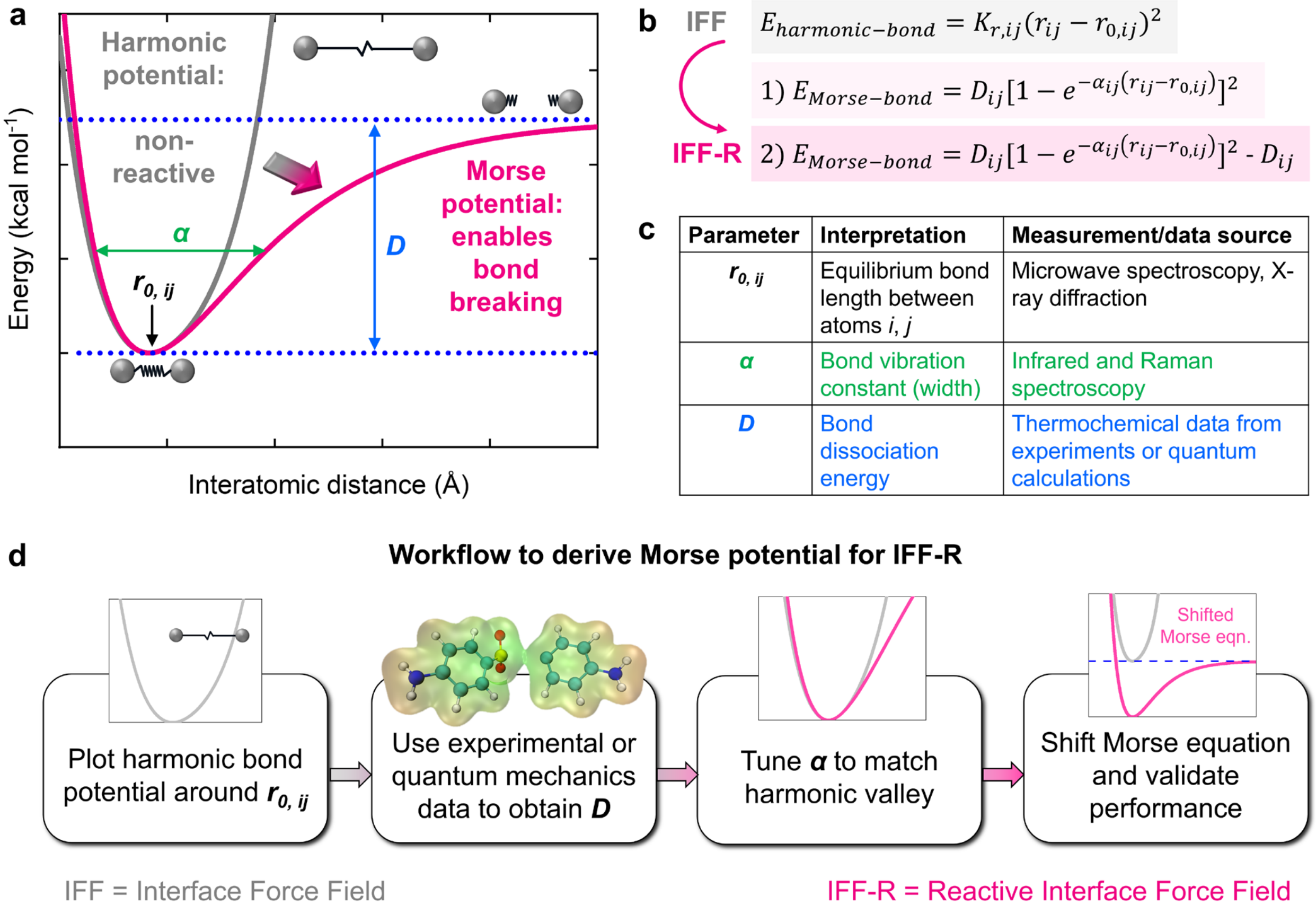

| Parameter | Interpretation | Measurement/data source |
|---|---|---|
| $r_{0,ij}$ | Equilibrium bond length between atoms *i*, *j* | Microwave spectroscopy, X-ray diffraction |
| α | Bond vibration constant (width) | Infrared and Raman spectroscopy |
| D | Bond dissociation energy | Thermochemical data from experiments or quantum calculations |

**Figure 1.** Concept of IFF-R parameterization. (a) The non-reactive harmonic bond potential in IFF (gray) is substituted for a dissociative Morse bond potential (pink) to yield IFF-R, focusing on chemical bonds at the lower spectrum of bond strength, or on all atoms in the system. The graph shows the bond energy vs. bond distance for the resulting Morse potential and the equivalent harmonic bond potential, as well as a bead-spring model for the resting state (at $r_0$), the stretched state (upper middle), and dissociated bond states, respectively (far right). The Morse potential in IFF-R is tuned to fit the harmonic bond potential near equilibrium ($\alpha$) and the bond dissociation energy ($D$) without changes in other, not bond-related force field parameters. The same modifications can be applied to other harmonic force fields (CHARMM, AMBER, OPLS-AA, DREIDING, CVFF, PCFF), given a certain level of thermodynamic consistency and interpretability of the parameters. (b) Changes in mathematical form and parameters for the harmonic and Morse bond potentials. Two parameters $r_{0,ij}$ and $K_{r,ij}$ from the harmonic parent force field are extended into three parameters in the Morse potential, of which all can be obtained from experimental data. If $r_{0,ij}$ and $K_{r,ij}$ are known from a parent force field, only $D_{ij}$ needs to be determined using additional information from experimental databases[49-52] or from quantum mechanical calculations, preferably at the MP2, CCSD, or higher level.[53] The Morse bond equation 1 is used for parameterization purposes only and the Morse bond equation 2 is strictly to be used in simulations to ensure a transition to zero bond energy during dissociation at larger atomic displacements. (c) List of parameters, interpretation, and data sources. (d) Workflow diagram to convert IFF into IFF-R using Morse potentials for specific chemistries.

**Results**

In the following, we describe the underlying theory, the workflow of parameter additions, examples of bond-breaking and stress-strain curves up to failure for validation, and the incorporation of bond forming reactions. The examples cover bond dissociation curves for small molecules, their representation in IFF-R, the prediction of stress-strain curves up to failure for multiple materials classes including carbon nanotubes (SWCNT), syndiotactic polyacrylonitrile (PAN), cellulose Iβ, spider silk protein, polymer/ceramic composites, and γ-iron. The computational speed is at least 30x higher than current state-of-the-art bond-order potentials. We also verify that IFF-R maintains the high-level performance for bulk and interfacial properties characteristic of IFF parameters, including validation of mass densities, vaporization energies, interfacial energies, and elastic moduli. We compare these critical computed pre-reaction and post-reaction properties between IFF-R and ReaxFF. We share an initial graphical user interface to aid in input preparation and simulation setup. We discuss limitations and potential extensions of IFF-

R. Hereby, our contribution focuses on the introduction of the method and providing sufficient evidence that the described IFF to IFF-R conversion can be applied to any carefully parameterized class I or class II force field to facilitate accurate simulations of bond breaking and formation. A large database of use cases, exhaustive lists of parameters, and optimizations of code are beyond the scope of this work.

**Description of the Approach.** The simulation of chemical reactions, especially in heterogeneous materials, has remained a major challenge in molecular dynamics simulations. The replacement of harmonic chemical bonds by Morse bonds (Figure 1a-c) in a simulation offers a simple and interpretable description of bond dissociation without the need for complex fit parameters, and has not been comprehensively realized in reactive force fields to-date.[54-56] To obtain the Morse parameters for a specific reactive bond, i.e., a pair of covalently bonded atom types *ij*, we can first plot the bond energy vs atomic distance using the harmonic bond potential, as known from experiment or the nonreactive potential (Figure 1a, d). Next, using experimental data[49-52] or high-level quantum mechanical estimates (CCSD(T), MP2, possibly DFT)[53] for the respective bond dissociation energy $D_{ij}$ (Figure 1c, d), we plot an initial Morse potential curve (Figure 1b) for atoms *i* and *j*. Thereby, the equilibrium bond length $r_{o,ij}$ remains the same as in the harmonic potential, or as known from experiment. The parameter $\alpha_{ij}$ can be used to fit the Morse bond curve to the harmonic bond curve near the resting state (Figure 1c, d) and typically in a narrow range of 2.1±0.3 $Å^{-1}$.[21] The parameter $\alpha_{ij}$ can be refined to match the wavenumber of bond vibrations in the simulations to experimental data from Infrared and Raman spectroscopy.[57]

Typical parameter ranges are 0.5 to 2.5 Å for equilibrium bond lengths, 1.8 to 2.4 $Å^{-1}$ for the width α (whereby larger values are equal to higher vibrational wavenumbers and a smaller width), and 0 to 250 kcal/mol for the bond dissociation energy. Examples of Morse bond parameters are shown in Figure 2 and in Table S1 in the Supplementary Information. The remaining force field parameters in IFF or other force fields (e.g., PCFF used here for some organic molecules, AMBER, CHARMM) can be retained without changes, including the atomic charges, Lennard-Jones parameters, angles, torsions, out-of-plane terms, and other non-reactive harmonic bonds. For simplicity of parameterization, Python code is provided in the Supplementary Software (see Figure S1 in the Supplementary Information for an overview) so that any available bond in IFF (or other class I and class II force fields) can be replaced with a Morse bond potential. The provided code performs an empirical fitting of the Morse bond potential to a given harmonic potential and may not represent the same level of IFF accuracy that will be discussed in the following subsections.

Further validation and quantum mechanical simulations indicate that the dissociation of reactive bonds can be considered complete between approximately 140% and 170% of the equilibrium bond length (Figures S2 to S6 and Section S3 in the Supplementary Information). We recommend removing topological connections between atoms *i* and *j* at approximately 200% of the equilibrium bond length to minimize artificial correlations in the simulation near and past failure. Following bond disconnection, different atomic charges and atom types may be assigned at the reaction site to maintain charge neutrality and accurately represent the reaction products.

**Morse Parameterization and Customization of IFF-R.** Bond dissociation curves with accurate values for the bond dissociation energy *D* are essential for reliable predictions of bond breaking mechanisms and mechanical properties in complex materials. Hereby, tabulated "bond dissociation energies" for specific bonds from the literature and available databases[50] are most suitable for IFF-R. "Average bond dissociation energies" are less suitable as they represent an average bond dissociation energy for all equivalent bonds in a molecule, which typically differ from the energy required to break the first bond, second bond, and so forth (e.g. multiple C-H bonds in $CH_4$).

As an example, we compared bond dissociation energies obtained using different methods for diethyltoluenediamine (DETDA) (Figure 2a, b) and 4,4'-diaminodiphenylsulfone (4,4'-DDS) (Figure 2c-f). The molecules are used in the synthesis of polyurethanes, polyimides and epoxy resins to manufacture high-strength CNT/epoxy composites.[58-60] We compare data from experiments (horizontal lines), 2$^{nd}$ order Møller-Plesset perturbation theory (MP2), density functional theory (DFT) with B3LYP, molecular dynamics simulation with tuned curves in IFF-R and ReaxFF (parameters from ref. [61]). The bond scans generally look similar but differ in detail. While experimental data are often an average for a given type of chemical bond, MP2 calculations resolve substituent effects and are more accurate than plane wave DFT methods. As a limitation, energy profiles at the DFT and MP2 levels invert toward lower energies at bond distances of 2.5 to 3 Å, as opposed to convergence to zero energy (Figure 2a-d). The unphysical curve shape is related to electron delocalization error (see further examples in refs. [62,63]) and impacts the reliability of DFT and MP2 methods in all instances where bonds form or dissociate, especially in catalysis such as $CO_2$ reduction, water splitting, fuel cells, and nitrogen fixation. The Morse potential in IFF-R as well as higher-level quantum mechanical calculations such as CCSD(T), if computationally feasible,[63] overcome these errors. Bond scans by ReaxFF appear somewhat irregular and less aligned with experimental or quantum mechanical results. IFF-R curves were tuned to a combination of experimental, MP2 and DFT values and yield smooth curve shapes (Figure 2a-f).

Highly resolved quantum mechanical calculations such as MP2, or preferably CCSD(T), can be used to distinguish differences in bond dissociation energies relative to average values for $C_{ar}$- $C_{ar}$, C-N, and C-S bonds from experiments.[49-52,64] Substituent effects, such as inductive and mesomeric effects, can be incorporated into IFF-R by choosing $\alpha$ and $D$ parameters that match MP2 or other high-level results. For example, in DETDA, the bond dissociation energy of the 1, 6 and 3, 4 bonds (MP2 in Figure 2a) is ~10 kcal mol$^{-1}$ lower than that of the 2, 3 and 5, 6 bonds (MP2 in Figure 2b). Hereby, the 1, 6 bonds and 3, 4 bonds have less double bond character compared to the 2, 3 and 5, 6 bonds according to an analysis of the resonance structures (Figure S7 and Section S4 in the Supplementary Information). In 4,4'-DDS, a push-pull effect by the $NH_2$ and $SO_2$ groups yields nearly full double bond character for the 2, 3 and 5, 6 bonds (MP2 in Figure 2c), which have a ~30 kcal mol$^{-1}$ stronger bond dissociation energy than the 1, 2 and 4, 5 bonds (MP2 in Figure 2d).[65] We did not incorporate specific adjustments in atom types and bond energies for substituent effects here, however, the data illustrate how IFF-R can include fine detail of the electronic structure when necessary. The MP2 bond scans also reproduce different bond strengths for $C_{ar}$-S and $C_{ar}$-N bonds in comparison to $C_{ar}$-$C_{ar}$ bonds, which are 75-85 kcal mol$^{-1}$, 110 kcal/mol, and 120-170 kcal mol$^{-1}$, respectively, according to experiments (Figure 2e, f).

In summary, IFF-R can consider all chemical bonds as breakable or focus on a sub-set of bonds, such as bonds of lowest strength only, and represent the remaining bonds as harmonic. All three Morse parameters for each bond type can be conveniently obtained using extensive databases with thousands of experimental measurements (Table S1 in the Supplementary Information), as well as using chemical analogy for similar bonds.[49-52,64] IFF-R accuracy can be further improved by tuning parameters based on observations from high-level quantum mechanical calculations. Thereby, IFF-R can account for specific chemical environments and electronic structure effects as needed.

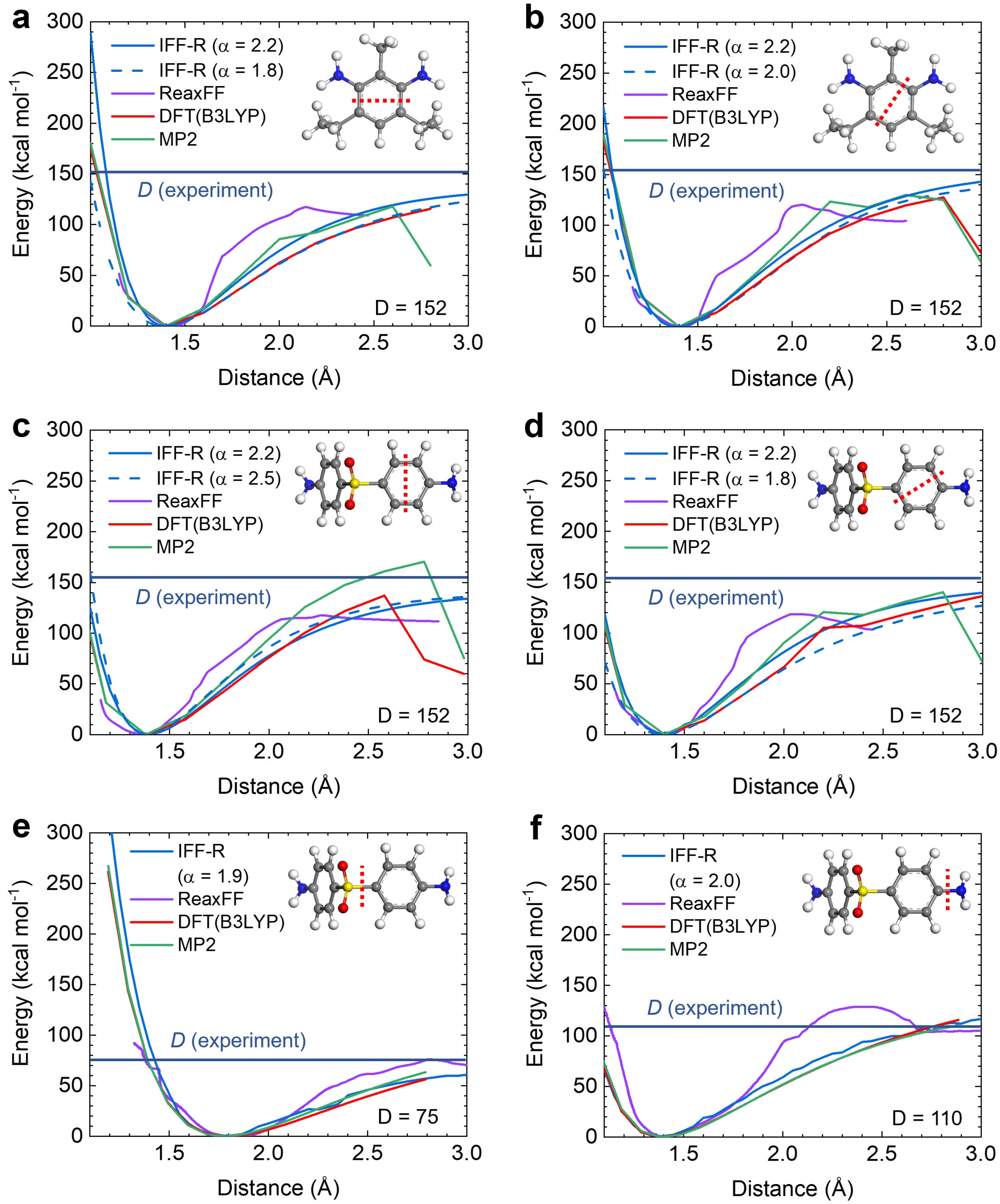

**Figure 2.** Examples of bond scans and experimental bond dissociation energies for the parameterization of the reactive INTERFACE Force Field, IFF-R. We consider the dissociation of several bonds and bond pairs in two amine monomers, 3,5-diethyl-toluene-1,6-diamine (DETDA) (a, b) and 4,4'-diaminodiphenylsulfone (4,4'-DDS) (c-f), indicated by red dashed lines. The experimentally determined bond dissociation energy *D* for similar chemistry is shown as a horizontal line and labeled in the bottom right corner of the panels (refs. [49-52,64]). These values, however, are somewhat approximate and do not consider the exact substitution pattern. Chemically specific bond scans were obtained using simulations at the level of 2nd order Møller-Plesset perturbation theory (MP2), density functional theory (DFT) with the functional B3LYP, and simulations with ReaxFF (ref. [61]). The drop in bond energies between 2.5 and 3 Å when using MP2 and DFT calculations is unphysical, related to electron delocalization errors (refs. [62,63]). IFF-R parameters were informed by experimental, MP2 and DFT data, and two options with different values for the bond vibration constant α are shown (solid and dashed blue lines). (a, b) Bond dissociation curves for $C_{ar}$ – $C_{ar}$ in DETDA in two distinct molecular planes. (c, d) Bond dissociation curve of $C_{ar}$ – $C_{ar}$ bonds in 4-4'-DDS in two distinct planes. (e, f) Bond dissociation curve of $C_{ar}$ – S and $C_{ar}$ – N bonds in 4,4'-DDS.

**Applications of IFF-R to Compute Stress-Strain Curves and Analyze Failure.** Stress-strain curves are critical to characterize mechanical properties and were simulated for different materials classes using IFF-R, including single wall carbon nanotubes (SWCNTs), polyacrylonitrile (PAN), cellulose Iβ, and γ-iron (Figure 3). Atomic-level deformations and the occurrence of bond scission upon failure can be clearly seen, which non-reactive simulations cannot predict. The comparison to experimentally measured Young's modulus and tensile strength,[66-70] or high-level quantum mechanical results if not available, shows good agreement (Figure 3a, b, d, f, h and Table 1).[3,67,68,71-74] The computed mechanical properties from IFF-R agree with most reference data within 10%.

The reliability is, on average, notably better and more consistent compared to ReaxFF, where typical deviations are 10% to 50% and exceed 1000% for iron (Figure 3 and Table 1). ReaxFF required the use of multiple parameter sets.[75-77] The choice of the ReaxFF parameter set is somewhat arbitrary for this study as the parameter sets were not explicitly parameterized for the purpose of uniaxial tensile simulations, and in some cases not even for the specific materials of interest. Therefore, the accuracy varied widely (Figure 3b, d, f, h). Given the difficulty of parameterizing ReaxFF parameters, it was necessary to use discretion to choose the best available ReaxFF parameters as a comparison. For convenience to the user, IFF-R and IFF, use one well-defined, interpretable parameter set that avoids ambiguity among multiple parameter sets, and overall thermodynamic consistency does not require specific fitting to compute mechanical properties.[44,61,75-81] Every compound, or functional group, can utilize dedicated atom types of the same element as necessary.

Regarding failure mechanisms, single wall carbon nanotubes (SWCNTs) often showed smooth cleavage planes (Figure 3a-c). The SWCNT fails by strain-induced bond dissociation propagation and reaches a strain of ~20% at break in the absence of defects (Figure 3a, c). Mechanical failure of the polymers PAN and cellulose Iβ arises from breaking of load-bearing bonds in single polymer chains (Figure 3d-g). Dissociation initially affects single chains, involves chain sliding, and then propagates further through the crystal (Figure 3e, g). The IFF-R model of γ-iron shows the evolution of microcracks and plastic deformation before failure (Figure 3h, i). Slip and crack propagation along partially preserved crystal planes is seen, leading to moderately ductile failure.

IFF-R produces continuous, non-linear stress-strain curves and reliable mechanical properties. The Young's modulus and tensile strength values for the SWCNT are comparable to experiments of a CNT with similar dimensions (Figure 3a, b).[68] For 100% crystalline PAN (Figure 3d-e), we found no experimental reference data for the tensile strength, and the elastic modulus was estimated to be 172 GPa from quantum mechanical DFT (GGA PBE) simulations using the Cambridge Serial Total Energy Package (CASTEP) program, which agrees with IFF-R data (Table 1).[82] ReaxFF with both the Liu and Damirchi parameters (refs. [75,77]) suggests similar stress-strain properties, although greater deviation from the DFT (GGA PBE) values are seen. The modulus of cellulose was computed about 20% lower than in experiment and the strength was overestimated by 20%, which is in better agreement and less ambiguous than the corresponding ReaxFF predictions (Figure 3f, g and Table 1). Cellulose parameters were used as given in PCFF and modified to utilize the Morse potential following the IFF-R protocol (Figure 1). Updating PCFF parameters for cellulose for higher IFF and IFF-R quality may require the representation of stereo-electronic effects and updated nonbond parameters and was not further considered here.

ReaxFF parameters for the γ-iron model led to rather divergent results, including more than twice the experimental strength and extreme, unphysical ductility of 40%. For γ-iron and other metals, IFF-R does not require the definition of Morse bonds due to the simplicity of the Lennard-Jones (LJ) potential and is equal to IFF without modifications. For most chemistries, including metals and oxides,[10,17] Young's Modulus tends to be better predicted using IFF and IFF-R with 12-6 LJ options (same energy expression as CVFF, AMBER, CHARMM, OPLS-AA, DREIDING) than with equivalent 9-6 LJ parameters (same energy expression as CFF, PCFF, and COMPASS).

We further tested the simulations of stress-strain curves up to failure with IFF-R for less idealized, more realistic nanostructures and biopolymers (Figures S8 and S9, and Section S5 in the Supplementary Information). These include SWCNTs with various levels of vacancy defects, which successively reduce the tensile strength and ultimate strain by ~20% (Figure S8 in the Supplementary Information). We simulated the failure of the spider silk protein (spidroin) at two different temperatures, demonstrating good agreement of the computed Young's modulus (3.6 GPa) and tensile strength (1.6 GPa) with experimental data (Figure S9 in the Supplementary Information). For detailed quantitative studies of the mechanics and failure mechanisms of complex materials, usually a large set of simulations using multiple morphologies, relaxation times, and strain rates is necessary.

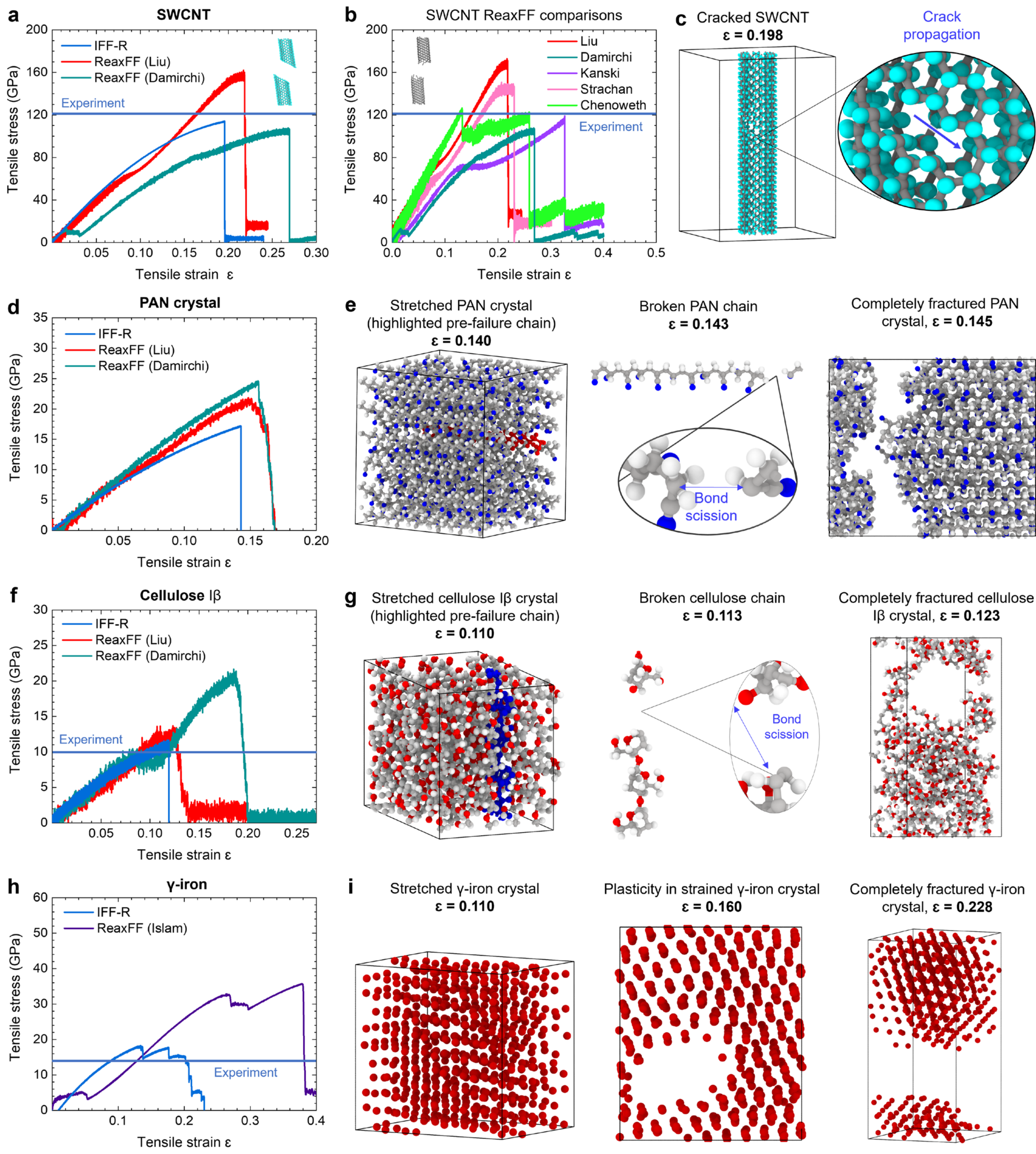

**Figure 3.** Stress-strain curves and snapshots of the materials under uniaxial tensile strain using the reactive INTERFACE force field (IFF-R), the reactive polymer consistent force field (PCFF-R), and ReaxFF parameter sets. When available, the experimentally determined tensile strengths, or high-level theory estimates, are shown as horizontal lines (a, b, d, f, h) (refs. [66-70]). (a, b) Single

wall carbon nanotubes (SWCNT) of 1 nm diameter and comparison to simulations with ReaxFF (refs. [44,75,77-79]). Models of the fractured CNTs are shown in the top right corner of (a) and top left corner of (b) (using ref. [77] in (b)). Virtual π-electrons are used for the IFF-R CNT model. (c) Crack propagation demonstrated in a SWCNT using IFF-R. (d) Syndiotactic, crystalline polyacrylonitrile (PAN). (e) Bond scission, progression of deformation, and failure in PAN. (f) Cellulose Iβ. (g) A broken chain in a strained cellulose Iβ crystal and evolution of cellulose Iβ failure. (h) Crystalline γ-iron. Strain was applied in the plane perpendicular to the (111) surface. (i) Plastic deformation and ductile failure in γ-iron.

**Table 1.** Young's Modulus and the tensile strength for SWCNTs, crystalline syndiotactic PAN, Iβ cellulose, and γ-iron (representative of steels) according to experimental measurements and molecular dynamics simulations with IFF-R, PCFF and ReaxFF (refs.[75-77]). The mechanical properties are reported in the alignment direction of CNTs and polymers, and the models are representative of different materials classes, including ceramics, polymers, biopolymers, and metals.

| | SWCNT | | PAN | | Cellulose Iβ (PCFF-R) | | γ-Iron, [111] direction[c] | |
|---|---|---|---|---|---|---|---|---|
| | Modulus (GPa) | Strength (GPa) | Modulus (GPa) | Strength (GPa) | Modulus (GPa) | Strength (GPa) | Modulus (GPa) | Strength (GPa) |
| Experiment | 1007 ± 118[a] | 121.6 ± 22[a] | 172 ± 10 (DFT) | N/A | 150 ± 5.0[83] | 4-10[66] | 210 ± 10[71] | 13.1 ± 1.5[67] |
| IFF-R (or PCFF-R) | 1001 ± 15 | 115 ± 1.5 | 168 ± 9.0 | 18.5 ± 2.0 | 120 ± 1.0 | 12.7 ± 1.5 | 237 ± 20 | 14.5 ± 1.1 |
| IFF-R dev to expt (%) | -1 ± 12 | -5 ± 19 | -2 ± 6 | NA | -20 ± 3 | +81 ± 43 | +13 ± 5 | +11 ± 11 |
| ReaxFF[b] | 752 ± 8.0 | 173 ± 2.0 | 219 ± 6.0 | 29 ± 2.0 | 131 ± 4.0 | 14.1 ± 2.0 | 4627 ± 12.0 | 35 ± 5.0 |
| ReaxFF dev to expt (%) | -25 ± 12 | +42 ± 19 | +27 ± 6 | NA | -13 ± 3 | +101 ± 43 | +2100 ± 5 | +169 ± 11 |

[a] From tensile testing of a 2.1 nm CNT using a micro/nanoscale material testing system (ref. [68]).

[b] Using the ReaxFF parameter set from Liu et al. (ref. [77]) for SWCNTs, PAN, cellulose and from Islam et al. (ref. [76]) for iron. Values may change for a different ReaxFF parameter set.

[c] Experimental data for γ-iron are similar to the mechanical properties of steel.[84,85] The [111] crystallographic direction was used, i.e., vertical to the (111) plane.

**Computational Efficiency.** A major benefit of simulations with IFF-R, besides accuracy and interpretability, is a high computational speed and low memory usage relative to alternative methods. The advantage over quantum mechanical simulations and DFT-machine learned potentials is obvious, about $10^6$ or $10^4$ times faster, respectively.[86] A comparison of the computational speed of IFF-R and ReaxFF using the open-source code LAMMPS shows on average about 30 times faster speed for systems sized between 1,000 and 10,000 atoms and needs only about 10% of the memory (RAM) (Figure 4). Therefore, IFF-R allows access to systems of larger size and more extensive dynamics. To arrive at this benchmark, we measured the computation time for the above stress-strain curves using molecular dynamics simulation in the NPT ensemble, normalized to 1 CPU with a typical time steps of 1 fs for IFF-R and 0.5 fs for ReaxFF. IFF-R is approximately 60x faster for simulations of metals, about 26x faster for small polymeric systems, and 8x faster for carbon nanotubes, which include virtual π electrons, when compared to simulations with ReaxFF (Figure 4).[75] IFF-R is faster by about the same magnitude for larger systems with over 100,000 atoms (Figure S10 and Section S6 in the Supplementary Information). The reason for the speed up lies in the simpler energy expression with an order of magnitude fewer terms, including constant atomic charges, predefined bonded terms, and approximately two orders of magnitude fewer algorithmic operations per unit time overall.

For the comparisons, it is important that exact numbers depend on the simulation code and settings of both IFF-R and ReaxFF. For example, 1 fs is a conservative time step for IFF-R and time steps of 1.5 fs or 2 fs can often be used along with the "shake" algorithm for H atoms and virtual π electrons.[87] The average speedup can then double to approximately 50x. Simulations can further be accelerated by using simulation codes other than LAMMPS, which is one of the slower open-source codes for both IFF/IFF-R and ReaxFF. IFF/IFF-R is most efficiently used with the open source code GROMACS, which runs 3 to 6 times faster than LAMMPS.[88,89] ReaxFF runs multiple times faster with the Amsterdam Modeling Suite (AMS), which is available commercially. The efficiency of ReaxFF in LAMMPS is strongly dependent on its implementation.[90] In this study, the "reax/c" LAMMPS implementation was used which is faster than "reax".[91] A major bottleneck in computational speed of ReaxFF is the handling of atomic charges.[77] This study used the "fix qeq/reax" method, however, some faster alternatives may exist and are not tested here (e.g., "fix qeq/fire").[92,93] The relative speedup can also be affected by use of GPUs rather than CPUs.

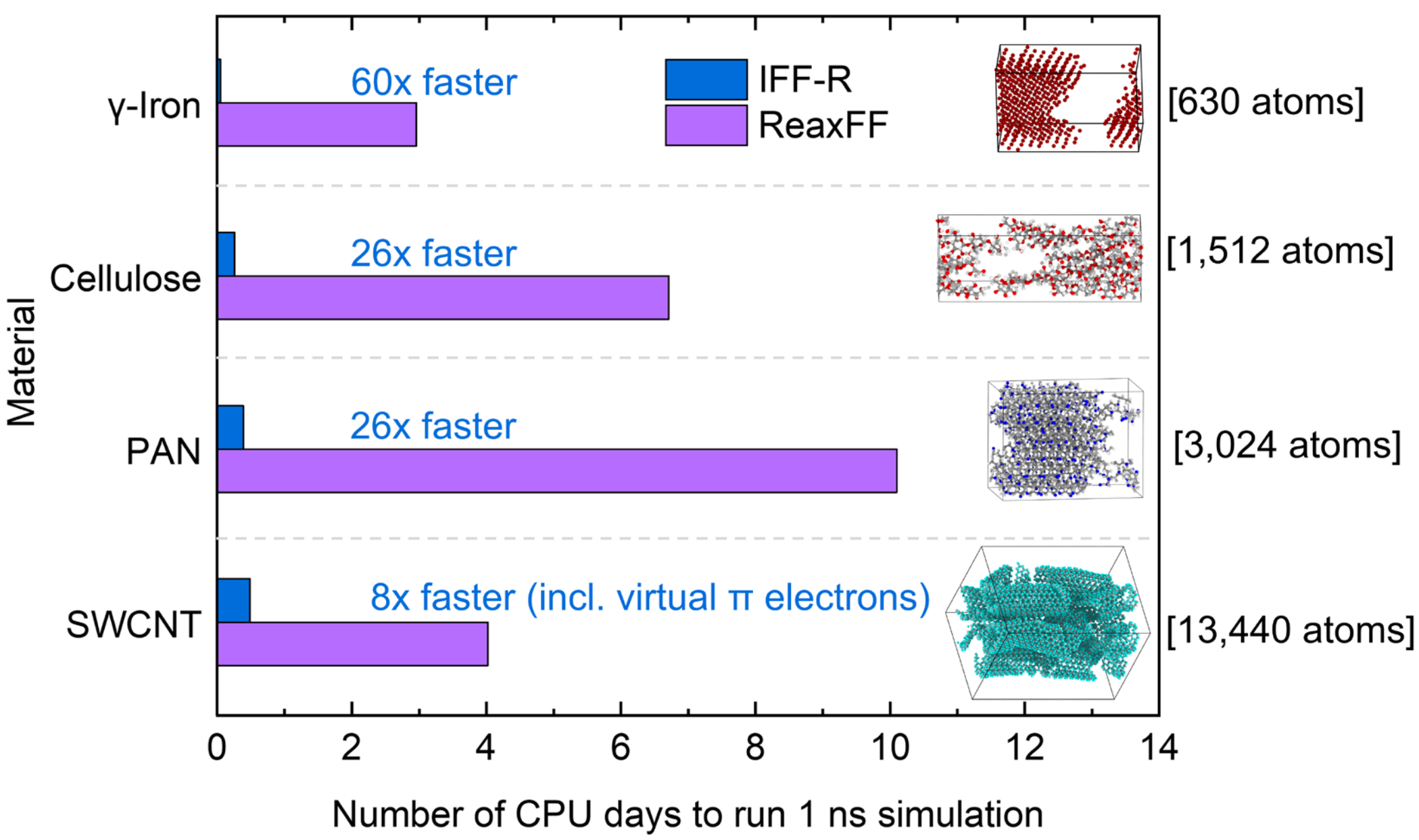

**Figure 4.** Comparison of computing time (CPU time) needed for the simulation of stress-strain curves of different materials using the reactive INTERFACE force field (IFF-R), or reactive polymer consistent force field (PCFF-R) in case of cellulose, and ReaxFF. The models were 600 to 14,000 atoms in size and include γ-iron, crystalline cellulose Iβ (ref. [94]), crystalline syndiotactic poly(acrylonitrile) (PAN), and a bundle of single-walled carbon nanotubes (SWCNTs) of 1 nm

diameter (which includes virtual π electrons in IFF-R). Snapshots at failure are shown and the speed-up in multiples of CPU time by IFF-R is indicated. The benchmark simulations were carried out on 1 CPU, using 1000 steps of molecular dynamics simulations with a constant strain rate in the tensile direction. The isobaric-isothermal ensemble (NPT) was employed with pressure control in $x$ and $y$ directions, and no pressure control in the tensile ($z$) direction. IFF-R completes 1 ns molecular dynamics simulation on 1 CPU in less than 12 hours for all 4 materials while ReaxFF requires 2 to 4 days for γ-iron and SWCNTs, and between 6 days and 1½ weeks for (bio)polymers. The relative CPU times remain similar for model systems over 100,000 atoms (Figure S10 in the Supplementary Information).

**Compatibility and Transferability**. We determined that the replacement of the harmonic bond potential in IFF with the Morse bond potential in IFF-R maintains the accuracy of all physical, chemical, and interfacial properties of IFF,[5] while adding the desired bond dissociation capabilities (Table 2). The key functionality consists in the unique compatibility to examine simulate mixed inorganic, organic, biomolecular, composite, and electrolyte model systems without additional fit parameters. Consistent with IFF theory,[5] validation of such functionality includes (1) the evaluation of correct structural properties via lattice parameters and the mass density of individual compounds, as well as (2) the evaluation of relative energy differences via surface or vaporization energies, both at 298 K. (3) In addition, compatible models require physically correct, interpretable representations of chemical bonding, which achieved by using the same atomic charges in IFF and IFF-R, and physically realistic Morse potentials. Altogether, keeping the same accuracy of predicted properties before and after the replacement of the harmonic potential with a Morse potential ensures that the key functions of the Hamiltonian are maintained, which is to predict structures and associated changes in energies.[5]

Specifically, the predicted lattice parameters, mass density, and surface energy of graphite using IFF-R remain the same as in IFF within statistical errors of ±0.5% (Table 2). The calculated lattice parameters and the surface energy are close to experimental data, and nearly equivalent for IFF and IFF-R. The same is true for the mass density and vaporization enthalpy of propionitrile, a small molecule most closely related to PAN, which is liquid at 298 K as opposed to solid graphite. The computed mass densities are within 1% of experimental data and comparable between IFF and IFF-R (Table 2). The simulated vaporization enthalpies of propionitrile are a few percent higher than experiment yet remain the same using IFF and IFF-R within the statistical uncertainty. α-D-glucose, the monomer of cellulose, shows nearly identical properties using IFF and IFF-R (Table 2). γ-iron shows identical performance for IFF and IFF-R, where covalent bonds were not necessary to facilitate reactivity (Table 2).

For α-D-glucose, we used "PCFF-R" parameters that correspond to PCFF parameters with addition of a Morse potential for the weakest C-O bonds, analogous to the extension of IFF parameters to IFF-R (Figure S11 in the Supplementary Information). PCFF and PCFF-R are of much lower accuracy than IFF, even though we included some improvements in the distribution of atomic charges (Table 2). The lattice parameters and vaporization (sublimation) energies computed by PCFF-R and PCFF match, and systematic errors relative to experimental data remain the same.[95] In terms of absolute values, the lattice parameters *a* and *b* exhibit 10% to 15% shrinkage and the *c* parameter expands by 36%, which is far from IFF standards (<0.5%),[5] and only the mass density and vaporization energy are comparable to experiment. Better force field parameters for glucose require extensive analysis and changes, such as stereo-electronic effects at C1 carbon atoms and more interpretable bonded and nonbonded parameters (Figure S11 in the Supplementary Information).

The agreement in 3D structures and relative energies of diverse compounds in IFF and IFF-R, as well as in PCFF and PCFF-R, confirms that a modification of harmonic bonds to Morse bonds maintains original accuracy and is efficient to introduce reactivity. The IFF-R protocol can also introduce reactivity into other force fields that use the same energy expression as IFF, e.g., CHARMM, AMBER, OPLS-AA, DREIDING, and PCFF.[5]

**Table 2.** Preservation of key properties upon changing the force field from IFF to IFF-R. Lattice parameters, mass densities, and surface energies (or vaporization enthalpies) from experimental measurements are compared to results from molecular dynamics simulations with IFF and IFF-R under standard conditions. For α-D-glucose, PCFF/PCFF-R parameters were utilized.

| Material | | Lattice parameters | | Mass density (g cm$^{-3}$) | Surface energy (S) (mJ/m$^2$) or vaporization enthalpy (V) (kJ mol$^{-1}$) |
|---|---|---|---|---|---|
| | | a, b, c (nm) | α, β, γ (°) | | |
| Graphite | Expt[96-98] | 0.246, 0.426, 0.680 | 90, 90, 90 | 2.26 | 186 ±2.0 (S) |
| | IFF | 0.247, 0.428, 0.675 | 90, 90, 90 | 2.24 | 188 ± 2.0 |
| | IFF-R | 0.247, 0.428, 0.676 | 90, 90, 90 | 2.23 | 188 ± 1.9 |
| Propionitrile | Expt[49,52] | N/A (liquid at 298 K) | | 0.782 | 36.1 ±1.6 (V) |
| | IFF | | | 0.786 | 37.7 ± 0.15 |
| | IFF-R | | | 0.783 | 37.0 ± 0.35 |
| α-D-glucose | Expt[95,99] | 1.036, 1.483, 0.493 | 90, 90, 90 | 1.56 | 181 ±5.0 (V) |
| | PCFF | 0.912, 1.268, 0.670 | 90, 90, 90 | 1.54 | 176 ± 0.26 |
| | PCFF-R | 0.912, 1.267, 0.670 | 90, 90, 90 | 1.55 | 176 ± 0.33 |
| γ-iron[a] | Expt[100] | 3.566, 3.566, 3.566 | 90, 90, 90 | 8.18 | 2250 ±100 (S) |
| | IFF[10] | 3.587, 3.587, 3.587 | 90, 90, 90 | 8.03 | 2270 ± 7.0 |
| | IFF-R | 3.587, 3.587, 3.587 | 90, 90, 90 | 8.03 | 2270 ± 7.0 |

[a] Experimental lattice parameters and surface energies were extrapolated for 298 K from experimental data in ref. [100], see also ref. [10].

**Comparison of the Accuracy of IFF-R and ReaxFF.** IFF-R typically computes structural parameters within 0% to 1% deviation relative to experiments for validated compounds, and ReaxFF with deviations of 0% to 5%.[17] In addition, a Hamiltonian foremost aims to reproduce energy differences associated with structural differences under standard conditions, such as surface energies (cleavage energies) or vaporization energies under standard conditions.[5] Mechanical, thermal, and reactive properties can only be meaningfully validated after this condition is met as they are defined as higher order derivatives of the energy with respect to interatomic distances and temperature. These concepts of thermodynamic consistency are integral to IFF-R and are often neglected in other interatomic potentials, making them less suitable and less extensible for the simulation of reactions. For example, some force fields reproduce structural and mechanical properties while energy differences exceed 100% deviation from experimental observations.[7]

IFF-R reproduces the surface energies and vaporization energies for various materials with less than 5% deviation with experiment (Table 3). For graphite, virtual π-electrons are essential to simultaneously reproduce surface, wetting, and adsorption properties.[101,102] Deviations using ReaxFF exceed 30% for graphite and propionitrile (Table 3). Mechanical properties such as modulus and tensile strength deviate in the 10-20% range from experiment with IFF-R, compared to a 50% range and maximum deviations over 100% with ReaxFF (Table 1) (see detailed comparisons in Section S7 in the Supplementary Information).

IFF-R performs better overall than ReaxFF for validated compounds, works reliably with a single parameter set, and can be applied to any mixture of materials (Tables 1 and 3). Further validation of IFF-R for predicting the thermomechanical properties of amorphous epoxy, benzoxazine, PEEK, and polyamide resins has recently been reported by Odegard et al.[103-106]

**Table 3.** Comparison of surface energies and vaporization energies of representative materials (ceramic, metal, organic) from experimental measurements with computed values using IFF-R and ReaxFF. Examples include the surface energies of graphite, iron), the vaporization energy of a molecular liquid (propionitrile) and the sublimation energy of a molecular solid (α-D-glucose). The reliability to compute these energy-related properties scales with the reliability to predict adsorption, stability of mixtures, interfaces, and reactions.

| Material | Experiment | IFF-R | Dev. of IFF-R (%) | ReaxFF[a] | Dev. of ReaxFF (%) |
|---|---|---|---|---|---|
| | | Surface energy (mJ $m^{-2}$) | | | |
| Graphite | 186 ± 2.0[97,98] | 188 ± 1.9 | +1.1 | 242 ± 2.0 | +30.1 |
| γ-Iron [111] | 2250 ± 100[100] | 2270 ± 7.0 | +0.9 | 2376 ± 8.0 | +5.3 |
| | | Vaporization energy ( kJ $mol^{-1}$ ) | | | |
| Propionitrile | 36.1 ± 1.6[52] | 37.0 ± 0.35 | +2.5 | 47 ± 0.9 | +30 |
| α-D-glucose | 181.7 ± 5.0[b, 99] | 176 ± 0.33 | -3.1 | 169 ± 1.6 | -7.0 |

[a] ReaxFF parameters from Liu et al. for graphite, propionitrile, and α-D-glucose (ref. [77]). Alternative Damirchi parameters (ref. [75]) are likely less suited. ReaxFF parameters by Shin et al. were used to calculate the surface energy of iron (ref. [80]). The alternative Aryanpour et al. parameter set for Fe (ref. [81]) would incur a deviation of 200% from experimental reference data and was not chosen here. [b] The experimental value of the vaporization enthalpy was extrapolated for 298 K from an exponential fit of the integrated form of the Clausius-Clapeyron equation (see ref. [99]).

**Disconnection of Morse Bonds after Dissociation.** IFF-R is suited to examine bond breaking mechanisms and analyze stress-strain curves up to failure. The Morse potential alone, however, does not disconnect the specified bonds upon elongation. Therefore, we systematically tested the disconnection of Morse bonds upon elongation for a range of cutoff distances, as well as the role of the shifted Morse potential using the program LAMMPS (Figures S3 to S5 and Section S3 in the Supplementary Information).[58] As a result of our tests, we recommend a cutoff at a bond elongation of about 200% of the equilibrium bond length. Depending on the bond type, a minimum elongation of 140% to 170% was required for physically meaningful results, and cutoffs much beyond 200% of the equilibrium bond length (e.g., 300%) introduce artificial energy contributions from the remaining bonded and nonbonded terms. At approximately 200% of the equilibrium bond length, typical Morse bond parameters overcome most of the attraction toward the minimum bond force (Figure 2). An additive shift of bond-specific Morse potentials to zero at the bond cutoff distance supports energy continuity when Morse bonds are disconnected and reduces discontinuities in other energy contributions (Figure 1). Upon disconnection of bonds, atomic charges and atom types can be reassigned to those of the reaction products, whereby keeping charge neutrality at the reaction site is essential (Figure S2 to S6 and Section S3 in the Supplementary Information). When simulations extend beyond bond disconnection, it is also possible to adjust chemistries for follow-on reactions, e.g., using reaction templates (Figure S6 in the Supplementary Information).

In comparison, we recall that quantum calculations at DFT and MP2 levels lead to unphysical shapes of bond dissociation curves and have potentially serious errors at around 200% of the equilibrium bond length (Figure 2a-d).[62,63] Therefore, DFT and MP2 data on bond dissociation are not suitable to train IFF-R beyond estimates of bond dissociation energies. IFF-R, in addition to good interpretability, is also promising to outperform reactive simulations based on machine learning with DFT and MP2 data, which incorporate the flawed bond dissociation curves (Figure 2a-d). Trustworthy guidance for bond formation or dissociation from ab-initio methods requires a theory level of CCSD(T) or higher.

**Simulation of Complete Chemical Reactions Including Bond Formation, Bond Dissociation, and Specific Stoichiometries.** In a similar way, the reversible process of forming bonds was implemented as reactive groups dynamically approach each other during the simulation. The physical basis to decide on bond connections or the occurrence of multi-step chemical reactions can involve local screening functions below a threshold distance, followed by changes in force field types, bond connectivity, and energy minimization as described by Barr et al.[107] Similar concepts were implemented in the REACTER framework in LAMMPS, for example, to polymerize nylon-6,6' and polystyrene with up to 99% cross-linking.[48] These methods can also be used to simulate bond cleavage, however, IFF-R with a shifted Morse potential better reproduces the characteristic nonlinearity of the bond energy before failure (Figure 1), eliminates discontinuities in bond energy, requires less parameters and less user input. IFF-R is compatible with REACTER to handle bond formation and bond breaking in the same simulation (Figures 5 and 6).

As a demonstration, we carried out reactive MD simulations to dissociate carbon-carbon bonds in polyacrylonitrile (PAN) and subsequently allow bond formation via a hypothetical self-repair mechanism (Figure 3d, e and Figure 5). First, nine aligned PAN chains were exposed to strain during IFF-R MD simulation until failure (Figure 5a, b). Subsequently, the model system was compressed and the fragments were allowed to simultaneously bond using the REACTER toolkit (Figure 5c).[48] While the mechanism is oversimplified here (and can be customized for specificity), the simulation demonstrates that IFF-R and REACTER are compatible to simulate bond cleavage and bond formation in a continuous simulation. The methods can be used, for example, to simulate self-healing of polymers.

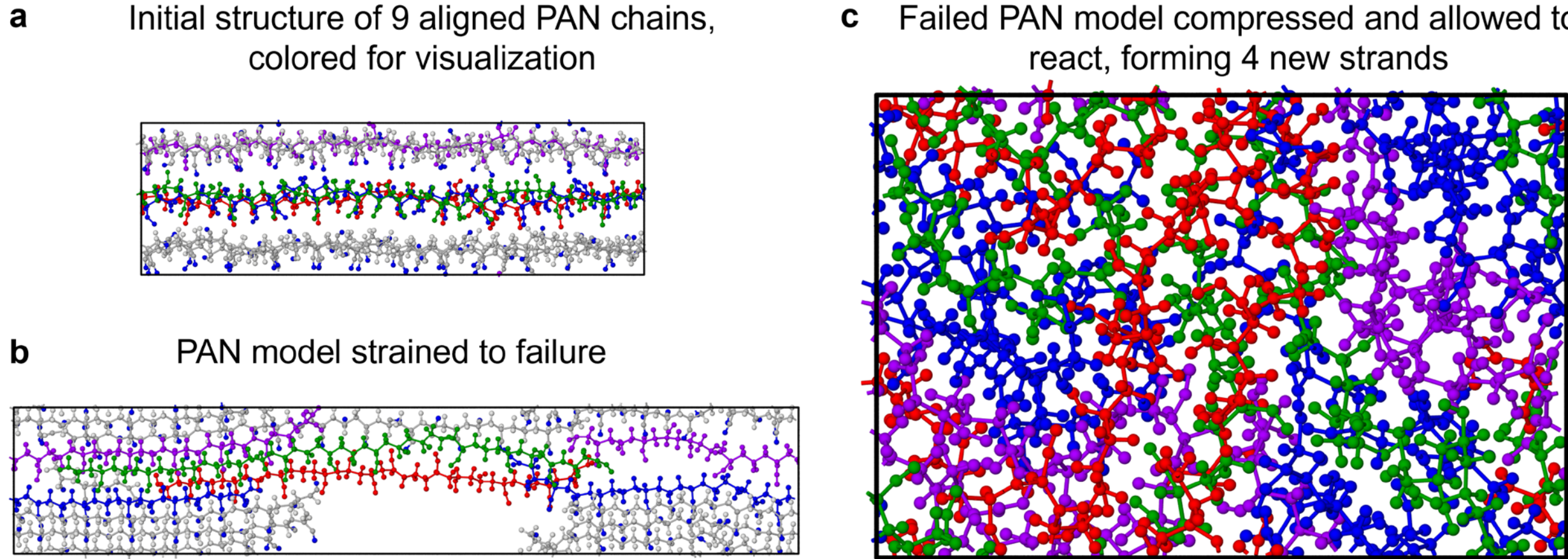

**Figure 5.** Simulation of micro-reversibility of bond breaking interactions. Tensile stress was applied to a model of aligned polyacrylonitrile (PAN) chains up to failure, followed by compression and allowing bond forming interactions using the reactive INTERFACE force field (IFF-R) and REACTER.[48] Bond breaking and bond formation occurred in a single, continuous simulation. (a) An initial, energy-minimized structure of 9 aligned PAN chains. 4 chains are shown in color to highlight the formation of 4 corresponding 4 PAN oligomers after chains failure and allowing reconnection of new bonds. (b) Bond failure during application of tensile stress in molecular dynamics simulation with IFF-R. (c) Compression of the disconnected structure from (b) and allowance of bond formation between broken chain ends using REACTER. Red, green, blue, and purple chains reacted with uncolored chains and formed 4 new polymer strands. A simplistic assumption of reversible bond formation was made for demonstration purposes. Specific mechanisms and statistical reactivity can be incorporated in detailed simulations.

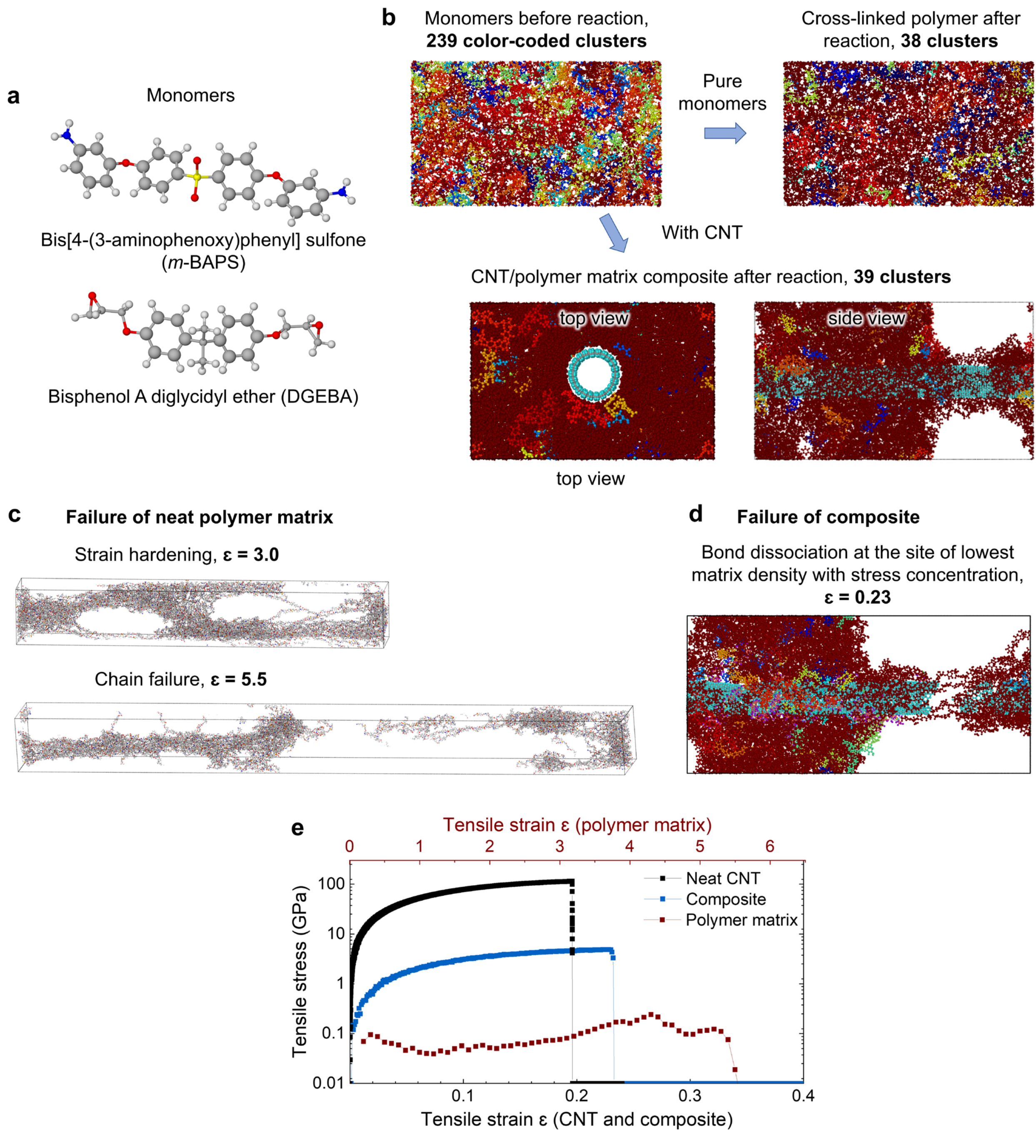

**Figure 6.** Molecular dynamics simulation of cross-linking an amine-epoxy network and a carbon nanotube (CNT)/epoxy polymer matrix composite, followed by the simulation of tensile strain $\varepsilon$ and failure mechanisms. (a) The two monomers bis[4-(3-aminophenoxy)phenyl] sulfone (*m*-BAPS), a difunctional amine, and bisphenol A diglycidyl ether (DGEBA), a difunctional epoxide,

were polymerized using the reactive INTERFACE force field (IFF-R) and REACTER.[48] The primary and secondary amine reactions were included. (b) A cluster analysis to visualize the number of independent molecules of the pre-reacted model (upper left), cross-linked neat polymer structures (upper right), and cross-linked polymer structures in the presence of a CNT (bottom left and right). (c) Snapshots of the cured epoxy resin in (b) during the strain simulation. (d) Snapshot of failure of the CNT/epoxy composite in (b). (e) Stress-strain curve (on logarithmic scale) for the cross-linked neat *m*-BAPS-DGEBA network, the CNT/epoxy composite, and a single CNT model.

As a second example, we simulated the polymerization and stress-induced failure of aerospace-grade epoxy polymers and a CNT/epoxy matrix composite (Figure 6). First, a stoichiometric mixture of epoxy and amine monomer precursors (Figure 6a) containing approximately 34,000 atoms was subjected to polymerization using IFF-R and REACTER[48] using the LAMMPS program (Figure 6b). Cross-linking simulations were carried out for the pristine monomers and for the monomers in the presence of a carbon nanotube. The polymerization reached ~84% cross-linking in both cases, whereby less than full conversion agrees with experimental observations due to limitations in diffusion of monomers and oligomers. Every molecule was color-coded and the progress of the reaction towards the cross-linked matrix can be seen as a color change (dark red color for cross-linked matrix molecules, due to superposition it may not appear quantitatively in the 2D view) (Figure 6b). Discontinuous matrix coverage near the top of the CNT reinforcement is seen in the side view. The models can be upscaled beyond $10^5$ atoms using IFF-R, which would be difficult using quantum calculations and bond order potentials. We used predefined reaction templates under standard pressure to form the cross-links (see Methods).

After the reactions reached completion, the matrix and composite models were subjected to MD simulations with an applied strain program until failure (Figure 6c-e). Stress-strain curves were recorded in single continuous simulations, showing clear differences in modulus, tensile strength, and strain at break. Failure mechanisms involved void formation, polymer-CNT bonding, and stress concentration in areas with minimal thickness of the matrix (Figure 6c, d). For the neat polymer matrix, the stress-strain curve shows an elastic region at low strain (0.0 to ~0.3), followed by yielding up to a strain of approximately 1.5, plastic strain hardening up to approximately 4.3, and failure with a strain at break of ~5.5 (Figure 6c, e). At a nominal strain of 3.0 (Figure 6c, top), the chains in the polymer network aligned with the strain direction (strain hardening) and at a nominal strain of 5.5, the polymer chains fully dissociated. The tensile strength was up to approximately 0.2 GPa, in the range of experimental data (Figure 6e). In the case of the CNT/epoxy composite, CNT reinforcement led to bond dissociation in the area with minimal matrix coverage and highest stress concentration (Figure 6d). The CNT/epoxy composite was stronger and less ductile than the polymer. Relative to the pristine CNT, the ductility was enhanced (strain at break 0.23) and the tensile strength decreased to ~6 GPa (Figure 6d, e). The pristine CNT had a tensile strength of 115 GPa and a strain at break of 0.20, consistent with laboratory data (Table 1).

Accordingly, IFF-R and other harmonic force fields can simulate bond formation and bond dissociation without the computational complexity of a bond order force field or extensive reparameterization, respectively. During the reaction phase, we recommend using the harmonic bond parameters to avoid unintentional dissociation of some atoms, which can induce failure of the simulation. Morse bond parameters should be available from the outset to run simulations continuously and applied once the reaction has reached the desired degree of conversion. To aid in this setup, we developed a tool for Morse bond parameterization and a sample script "in.xlink" (Section 2 in the Supplementary Information and Supplementary Software).

**Opportunities, Potential Pitfalls, and Limitations.** The proposed Morse bond modification can be applied to all force fields using a harmonic, or otherwise exponential bond potential, such as IFF, CHARMM, AMBER, PCFF, COMPASS, OPLS-AA, and DREIDING. When the parent force fields do not reproduce experimentally known properties such as lattice parameters or vaporization energies as seen for α-D-glucose and cellulose in PCFF (Table 2), we recommend a derivation of updated parameters following IFF protocol[5] to improve the accuracy and transferability, which is of equal importance to adding Morse potentials.

In the derivation of the three Morse parameters for new reactive bonds, we recommend to strictly follow the physical interpretation, which includes the equilibrium bond length ($r_{0ij}$), bond stiffness with associated vibration spectra ($\alpha$), and bond dissociation energies ($D$) (Figure 1). Experimental data as well as simulations with high-level quantum mechanical methods such as CCSD(T) and MP2 are suitable. Quantum mechanical simulations at simplified DFT level often have large errors and should be avoided for parameterization purposes.[108,109] Alternatively, Morse parameters could be empirically adjusted to match computed stress-strain properties to experimental reference data. We would give low priority to this approach as it may not correctly weigh contributions from other energy terms to the mechanical properties (angles, torsions, charges, LJ potential) and insert additional uncertainties from stress-strain measurements related to chemical composition and crystallinity of the samples into the force field.

Chemical reactions after bond dissociation can involve a variety of intermediates and products, including rearrangements with new chemical bonds, addition, elimination, and other multistep reactions (Figure S6 in the Supplementary Information). Morse bonds in IFF-R do not consider secondary reactions unless reaction templates with defined changes in bond connectivity are added. The creation of reaction templates remains specific for a given type of chemistry. Likely reaction mechanisms and their representation in molecular simulations need to be known *a priori* from chemistry knowledge, for example, simple one-step stoichiometric conversions to products, inclusion of intermediates, and multiple products. Parameters for reaction intermediates and products meeting IFF standards may be needed. Predictions of reaction mechanisms by computational methods such as quantum mechanics/DFT, ReaxFF, or IFF-R are hardly feasible since the details often depend on solvents, pH value, temperature, catalysts, byproducts, and time scale (e.g., seconds to hours).[65,110] Synthetic chemists continue to rely on experimentally derived knowledge, enhanced by insights from computational methods.

**Discussion**

IFF-R is well suited for the analysis of deformation and failure mechanisms of complex biological and material structures from atomic to micrometer scales. IFF-R builds on an effective replacement of nonreactive harmonic bond potentials with reactive Morse bond potentials in the Interface Force Field (IFF) and in other harmonic force fields to allow the simulation of bond scission. The modifications are easy to implement via substitution of the harmonic bond expressions with 2 interpretable parameters by a Morse-bond expression with 3 interpretable parameters, which are shifted to zero energy at the bond cutoff. We call this force field IFF-R and other harmonic reactive force fields equivalently, e.g., CHARMM-R and PCFF-R. Morse bonds can be assigned to selected bonded atom pairs or to all bonded atom pairs, and the remaining force field parameters can be used as is without loss in accuracy. Bond formation during chemical reactions can be incorporated by template-based methods such as the REACTER framework, enabling the dissociation and formation of bonds in single, continuous simulations using IFF-R.

IFF-R is suitable to simulate non-linear stress-strain curves up to failure and predicting elastic moduli and tensile strengths for all material classes and multiphase materials across the periodic table with high accuracy, at low computational cost, and with high interpretability. Relative to IFF, IFF-R improves the accuracy of the representation of chemical bonding and retains the accuracy of computed physical, chemical, and interfacial properties, including lattice parameters, mass densities, surface energies, and elastic moduli. The parameters for Morse bonds can be derived from experimental and high-accuracy ab-initio data and, as an option, also include local electronic structure effects via customized atom types with specific bond dissociation energies.

IFF-R uses a single parameter set, simplifying the user-experience and avoiding the challenge of selecting from a multitude of differently performing parameter sets, as is common with electronic density functionals and ReaxFF. The accuracy is up to several times higher than other reactive force fields, such as ReaxFF. The computational speed is about 30 times faster when using LAMMPS and up to 100 times faster using more efficient MD codes like GROMACS. IFF-R provides dependable predictions for deformations and failure of bulk materials, heterogeneous interfaces, and composite materials up to a billion atoms and length scales of a micrometer, supporting the generation of reliable large data sets for machine learning and accelerated multiscale simulations.

The representation of chemical reactions involving bond formation requires additional assumptions, e.g., reaction templates, and can be integrated using the REACTER framework. Hereby, reaction pathways and parameter sets for products need to be provided, or determined if necessary, before the simulation. The derivation of force field parameters remains the same as for IFF, the only difference lies in using a Morse potential for covalent-type bonds. A general on-the-fly parameterization of IFF-R would be desirable to further simplify reactive simulations for the user, however, such a method requires further work due to the complexity of physical principles and chemical analogy involved, case-specific search and curation of reliable reference data, and extensive simulations for validation.

## Methods

Molecular models were built using the Materials Studio 7.0 Graphical User Interface[111], and molecular dynamics simulations were carried out using the Large-scale Atomic/Molecular Massively Parallel Simulator (LAMMPS, version Sep2021),[58] as well as using the Discover program. OVITO[112] 2.8, VMD,[113] the Multiwfn[114] software, and Materials Studio were employed to visualize MD trajectories and prepare images of the simulation models. REACTER map files were created manually, and the AutoMapper tool developed by Bone et al.[115] was used for conversion of REACTER templates to LAMMPS molecule files.

*Bond scan simulations.* Ab-initio simulations of bond scans were performed using MP2 calculations based on Hartree-Fock theory with added electron correlations and DFT with the B3LYP gradient-corrected hybrid functional using Gaussian 2016. The amine monomers were aligned in the x-y plane of the simulation cell and the bond length was increased in increments of 0.1 Å or less. Thereby, atoms not participating in the bond were constrained in their relative positions. Energy minimization was performed at each bond length with $10^{-6}$ eV total energy convergence to construct the zero-point bond energy profile. Molecular mechanics simulations of bond scans with IFF-R and ReaxFF involved energy minimization using the conjugate gradient algorithm in LAMMPS for 100 to 500 steps (up to 10,000 steps for verification) with an energy tolerance of $10^{-4}$ and a force tolerance of $10^{-6}$ kcal mol$^{-1}$ Å$^{-1}$) (Section S1.1. in the Supplementary Information).

*Simulations of stress-strain curves.* For the simulations of stress-strain properties up to failure, models were initially subjected to energy minimization. Subsequent molecular dynamics simulations with IFF or IFF-R employed a temperature of 298.15 K, setting initial atomic velocities using a Gaussian distribution, and the Nose-Hoover thermostat for temperature control including dampening within 100 timesteps. We employed the NPT ensemble, except for CNTs that include some vacuum in the simulation box where the NVT ensemble is more suitable, and the Nose-Hoover barostat for pressure control at 1 atm. The tensile strain was increased continuously during molecular dynamics simulations in a strain program from zero until after failure. The maximum strain varied depending on the material from 0.4 for CNTs to 6.0 for polymers. The strain rate was between 0.04 and 0.10 per 100 ps, equal to between 0.4 ns$^{-1}$ and 1.0 ns$^{-1}$. All components of the stress tensor were measured using the NPT ensemble for polymers and metals,[116-118] and using the NVT ensemble for CNTs. In the case of CNTs, the engineering stress in the tensile direction was calculated using the cross-sectional area of the CNTs in relation to the total cross-sectional area of the box that included some vacuum (Section S1.2 in the Supplementary Information).

The summation of Coulomb interactions used the PPPM K-space solver, including a cutoff for the direct summation at 8.0 Å and relative energy tolerance of $10^{-4}$ for the long range, corresponding to high accuracy. The summation of Lennard-Jones interactions, which represent van-der-Waals interactions, was carried out with a spherical cutoff at 12.0 Å which is standard for IFF. The time step was typically 1 fs and lower at 0.5 fs for models with hydrogen or virtual π-electrons. Tensile simulations were used to strain periodic models along the axial direction of the carbon nanotube or polymer chains, and perpendicular to the [111] facet for the iron crystal at a strain rate of $10^{-6}$ every 10 timesteps with pressure control at 1 atm in the lateral directions. The simulations were run until 100% strain or material failure. Stress was monitored relative to the initial cross-sectional area of the models. The tensile modulus and the tensile strength were determined from the computed stress-strain curves, e.g., modulus as a ratio of stress over strain at low deformation (strain < 0.01). Standard deviations for the tensile modulus and strength were obtained from three independent simulated models with different initial velocities.

Additional simulations at the DFT level for PAN are described in Section S1.3 in the Supplementary Information.

*Evaluation of structural and surface properties.* The evaluation of equilibrium crystal structures and liquid densities also utilized molecular dynamics simulations in the NPT ensemble. The computation of surface energies (using equilibrium lateral dimensions of the surface from simulations in the NPT ensemble), simulations of stress-strain curves of CNTs, and of the average energy of gas molecules as part of the calculation of the vaporization energies utilized the NVT ensemble. The reason to use the NVT ensemble is that these simulation setups include major parts of vacuum or gas phase.

Equilibrium crystal structures and liquid densities were obtained by subjecting 3D periodic bulk structures to 1 ns molecular dynamics at 1 atm pressure and 298 K in the NPT ensemble. The average lattice parameters from this simulation, in case of crystalline materials graphite and iron, were further used to create models with cleaved surfaces by expanding the simulation cell 4.0 nm in the *z*-direction, creating an implicit surface because of the periodic z-boundary. Hereby, we used the (0001) cleavage plane for graphite and the (111) cleavage plane for iron. The cleaved structure and the bulk structure with average lattice dimensions were subjected to 1 ns of molecular dynamics at 298 K in the NVT ensemble. The average energies for the cleaved and bulk models were used to calculate surface energy, normalized relative to the surface area (Section S1.4 in the Supplementary Information).

*Evaluation of vaporization and sublimation enthalpies.* Calculations of the molar enthalpy of vaporization and sublimation ($\Delta H$) were performed by obtaining average energies for the crystalline (α-D-glucose) or liquid (propionitrile) structures, respectively, and for the same structures in the gaseous state, using the energy difference ($\Delta U$) and a correction for volume work ($\Delta H = \Delta U + PV$). Simulations of the crystalline and liquid models were run for 1 ns at 298 K and 1 atm pressure in the NPT ensemble. Simulations of the gases were carried out for 1 ns at 298 K in the NVT ensemble. The average energy per molecule was used to calculate the enthalpy of sublimation or vaporization as a difference between the gaseous and the solid or liquid state, respectively. Standard deviations for computed surface energies, enthalpies of vaporization, and enthalpies of sublimation were obtained from the block averages of the total energies in equilibrium (Section S1.5 in the Supplementary Information).

Examples for deriving IFF-R parameters and improving nonreactive base parameters (e.g., PCFF) are given in Sections S1.6 and S1.7 in the Supplementary Information.

*Simulation of bond-forming reactions.* We used the REACTER framework to facilitate bond formation.[48] The required reaction templates were created in Materials Studio and the molecules (PAN, amine/epoxy) were allowed to react during MD simulation at 1 atm pressure over a period of 150 ps using a reaction distance of 6.0 Å, which included equilibration and a reaction phase of 50 ps. Two thermostats were used during the reaction phase: (1) a Nose-Hoover thermostat at 600 K and 100 fs dampening for all atoms not participating in the reaction and (2) a velocity rescale thermostat at 200 K and 100 fs dampening for all atoms participating in the reaction. This setup uses time-temperature equivalence to accelerate the reaction time from milliseconds in real time to less than nanoseconds in the simulation. Atoms not participating in the reaction received enough kinetic energy at 600 K to move more rapidly and explore the range of potential reaction constellations in a short timeframe of 50 ps. During the reaction, atoms are relatively near to each other and may experience large forces upon connection, as well as strongly repulsive interactions when nonbonded at short distance or overlapping. Limiting the temperature to 200 K in the reacting regions mitigates such interactions and prevents failure of the simulations. Lower reaction distances, e.g., 4 Å to 5 Å, along with longer simulation times can also be explored and reduce the need for two dissimilar thermostats (Section S1.8 in the Supplementary Information).

The concept of IFF-R was originally described on the Heinz group website in January 2020 along with a tutorial[119] and an ArXiv preprint.[120] Supplementary Data 1 include sample force field files, simulation input scripts, 3D molecular models to reproduce simulations, , as well as a tutorial describing the workflow and examples for IFF-R. Supplementary Software contains code for automated Morse bond conversion and a new LAMMPS option for the shifted Morse potential. The Supplementary Information includes additional figures and discussion on the benefit of the shifted Morse potential versus the standard Morse potential, stress-strain curves and total energies up to failure for different bond cutoffs, individual energy contributions as a function of strain, reaction pathways after bond dissociation, the relationship between resonance structures and bond strength for selected examples, further mechanical property analysis for select materials, the comparison of the computational speed of IFF-R for large systems >100,000 atoms relative to ReaxFF, as well as details of the IFF and IFF-R parameterization of organic molecules.

Supplementary tables contain example parameters for the Morse potential and further comparisons of computed properties using IFF-R and ReaxFF to experimental data.

## Data Availability

All data to reproduce the results are available in the manuscript, Supplementary Information, Supplementary Data 1, and can also be obtained upon request from the corresponding author. Specifically, Supplementary Data 1 contains 3D atomic models, force field files, and simulation scripts to reproduce the results. Supplementary Software contains code to deploy the newly developed IFF-R tools, including examples. The same data and code are also shared on Zenodo.[121]

## Code Availability

All code is available in the Supplementary Software, including the GUI and the custom LAMMPS option for the shifted Morse potential. The same data and code are also shared on Zenodo.[121]

**Acknowledgements**

This work was supported by the NASA Space Technology Research Institute (STRI) for Ultra-Strong Composites by Computational Design (NNX17AJ32G), the National Science Foundation (CMMI 1940335, OAC 1931587, DMREF 2323546), BASF SE (Ludwigshafen, Germany), and the University of Colorado at Boulder. We acknowledge the allocation of computational resources at the Summit supercomputer, a joint effort of the University of Colorado Boulder and Colorado State University, which is supported by the National Science Foundation (ACI-1532235 and ACI-1532236), and resources at the Argonne Leadership Computing Facility, a DoE Office of Science User Facility supported under Contract DE-AC02-06CH11357.

**Author Contributions**

J. J. W. developed the workflow, carried out nonreactive and reactive MD simulations with IFF, IFF-R and ReaxFF, analyzed the data, and wrote the manuscript. K. K. contributed to the conception and workflow of the study, simulations, and proofreading. J. K. developed workflows, coded and tested the graphical user interface to assign Morse bonds in IFF-R, carried out simulations, and wrote the manuscript. P. J. i. V. coded the custom LAMMPS option for the IFF-R Morse potential, tested calculations, and proofread the manuscript. G. S. and R. P. carried out

MP2 and DFT calculations of bond scans. B. D. and A. v. D. carried out reactive simulations with ReaxFF and wrote the manuscript. G. M. O. advised J. K., wrote and revised the manuscript. H. H. conceived the study, coordinated the effort, analyzed the data, wrote, revised, and proofread the manuscript and supporting files.

## Competing Interests Statement

The authors declare no conflicts of interest.

**Supplementary Information**

for

# Implementing Reactivity in Molecular Dynamics Simulations with Harmonic Force Fields

by

Jordan J. Winetrout,[1,2] Krishan Kanhaiya,[1,2,3] Joshua Kemppainen,[4] Pieter J. in 't Veld,[5] Geeta Sachdeva,[6] Ravindra Pandey,[6] Behzad Damirchi,[7] Adri van Duin,[7] Gregory M. Odegard,[4] Hendrik Heinz[1,2*]

[1] Department of Chemical and Biological Engineering, University of Colorado at Boulder, Boulder, CO 80309, USA

[2] Materials Science and Engineering Program, University of Colorado at Boulder, Boulder, CO 80309, USA

[3] Insitute of Physics, Ruhr University Bochum, Universitätstrasse 150, 44801 Bochum, Germany

[4] Department of Mechanical Engineering - Engineering Mechanics, Michigan Technological University, Houghton, MI 49931, USA

[5] BASF SE, Molecular Modeling & Drug Discovery, 67056 Ludwigshafen, Germany

[6] Department of Physics, Michigan Technological University, Houghton, MI 49931, USA

[7] Department of Mechanical Engineering, Pennsylvania State University, University Park, PA 16802, USA

* Corresponding author: hendrik.heinz@colorado.edu

**This PDF file includes:**

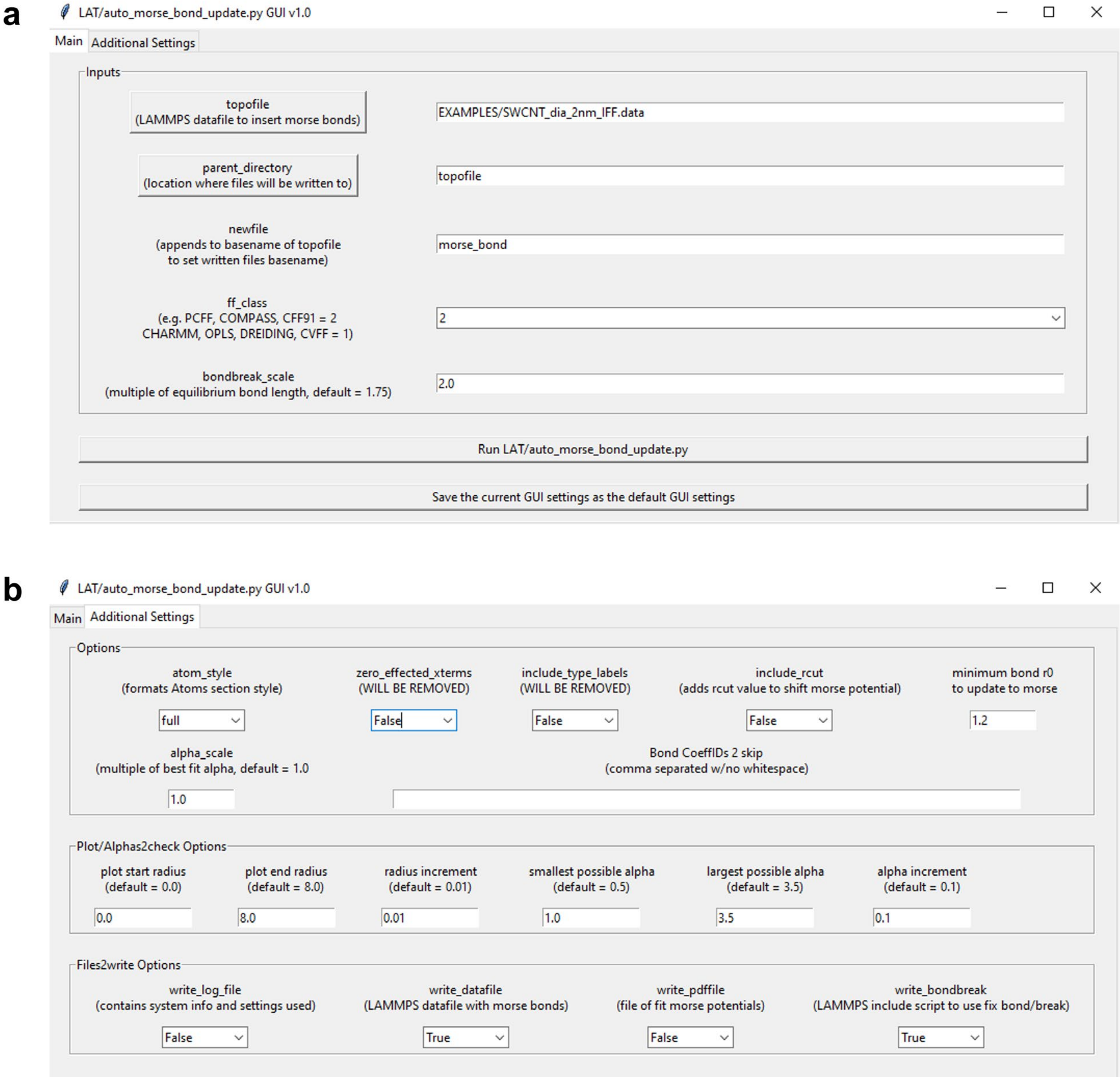

**Figure S1.** Snapshot of the graphical user interface (GUI) for Morse bond conversion from a classical, harmonic force field (auto_morse_bond_update.py). The input to the code is a LAMMPS data file with classical bond parameters. The outputs include a data file with Morse bond parameters, a LAMMPS input script file containing “fix bond/break” commands for the newly assigned Morse bonds, a pdf file showing Morse bond parameterization graphically, and a logfile with system information that can be useful for tracking parameter assignment. (a) The main tab

used for automatic Morse bond conversion. The "topofile" button is used for selecting the LAMMPS data file for Morse bond conversion. The "parent_directory" button is the default path that determines the directory for the outputs, and "topofile" is the path of the LAMMPS data file (topofile). If "topofile" is not used for the "parent_directory" option, the full path to the desired output directory should be given. The "newfile" section is an additional descriptor appended to the "topofile" name to identify the new data file with Morse bonds and avoids overwriting the "topofile" if it is in the same output directory. The auto_morse_bond_update.py code will work with class I and class II force fields, and the "ff_class" section needs to designate 1 or 2 as acceptable inputs, accordingly. The "bondbreak_scale" is used cutoff distances for the new Morse bonds (used with the "fix bond/break" command), and the unit is the multiple of the equilibrium bond length. We recommend this parameter to be 2.0, e.g., creating a bond/break cutoff for a C-C single bond with $r_0$ = 1.54 Å at 3.08 Å. (b) Image of the "Additional Settings" tab. This tab is included so the user has customizability over the outputs from the auto_morse_bond_update.py code. The "Options" section provides capabilities for the user to modify the information written to the output datafile. The "Plot/Alphas2check Options" section allows the user to change variables associated with the automatic parameterization of a Morse bond. Finally, the "Files2write Options" section allows the users to specify which files should appear in the output directory with "False" meaning the file will not be written to the output directory and "True" meaning the file will be written to the output directory. Using this menu of additional settings is optional.

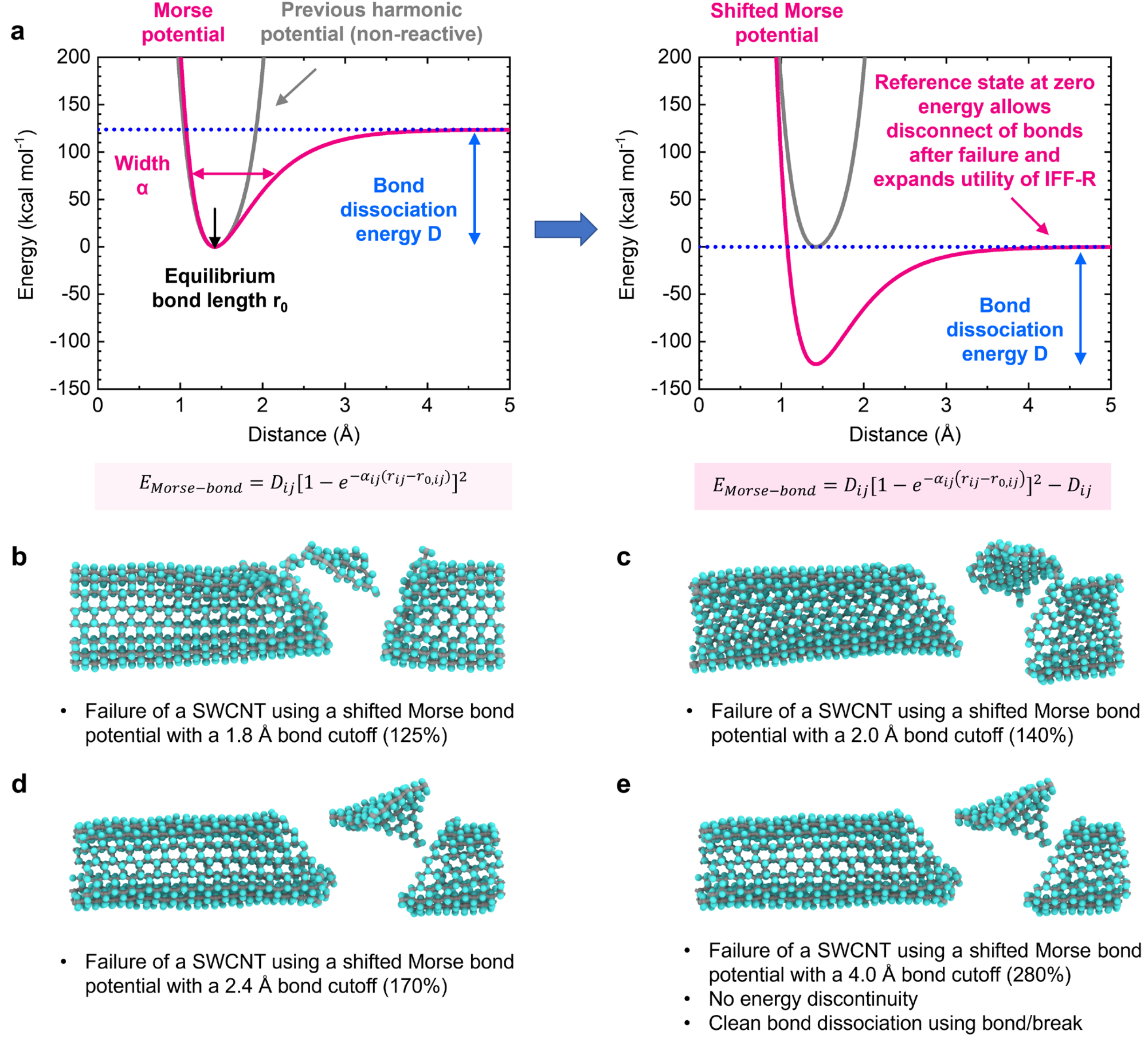

**Figure S2.** Illustration of the shifted Morse bond potential in IFF-R and implications on failure (see Supplementary Software "code for custom LAMMPS option user morse-2Aug23"). The Morse potential benefits from energy conservation, i.e., a shift to zero energy upon bond dissociation. (a) The default Morse bond potential (left) and the shifted Morse bond potential (right) for graphitic carbon-carbon bonds with an equilibrium bond length of 1.42 Å. The unmodified Morse bond potential has an energy minimum at $r_{0,\,ij}$ and 0 kcal mol$^{-1}$ and a high potential energy upon bond dissociation, which causes an energy discontinuity (energy drop) when elongated Morse bonds are disconnected or reassigned, respectively. In comparison, the shifted Morse bond potential has an energy minimum at $r_{0,\,ij}$ and $-D_{ij}$. Using a shift so that the bond energy approaches 0 kcal/mol as the bond dissociates avoids energy discontinuities in molecular dynamics simulations arising from the Morse potential during bond scission. At the same time, the shift does not change the force as a function of distance and has no visible effect on the failure mechanism. (b-e) Another key parameter is the bond cutoff distance. If the bonds are not disconnected at a certain distance past dissociation, "ghost" contributions from remaining angle, dihedral, and other energy contributions taint the potential energy. On the other hand, cutoff distances too close to the equilibrium bond length are unphysical. Examples of the fracture pattern of a SWCNT are shown using the shifted Morse potential with bond cutoff distances of 1.8 Å (b), 2.0 Å (c), 2.4 Å (d), and 4.0 Å (e). The simulation protocol was the same. The failure mechanism was affected when the cutoff distances were chosen too small and became independent of the cutoff distance when chosen larger than 2.4 Å, equal to ~170% of the equilibrium bond length.

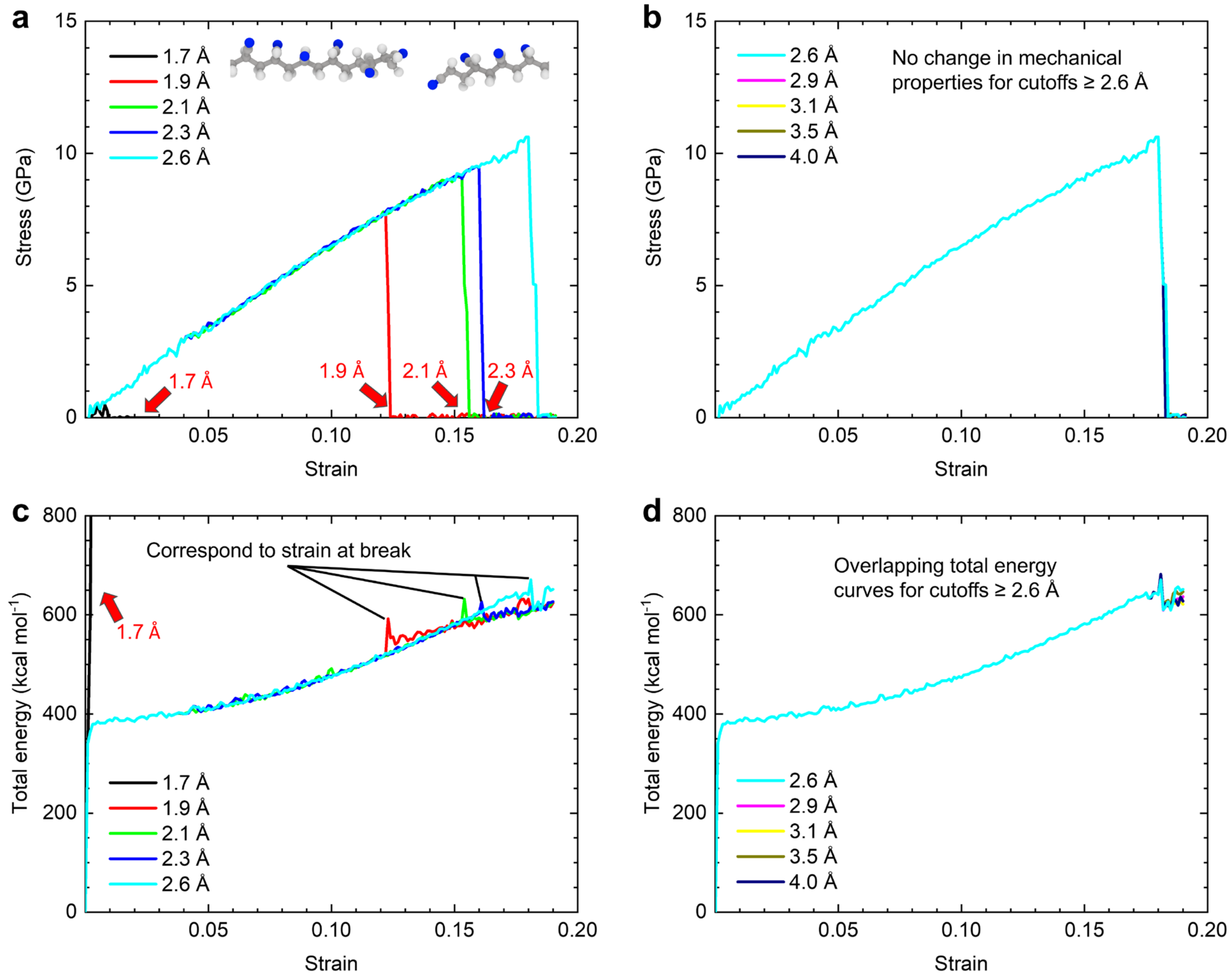

**Figure S3.** Stress-strain curves of a polyacrylonitrile chain and the total energy for 9 different bond cutoffs ($R_{max}$) using a Morse potential shifted by $-D_{ij}$. The bond/break LAMMPS command was used during a simulation of the tension response. (a) Stress-strain curve for $R_{max}$ values from of 1.7 Å to 2.6 Å, equal to 111% to 170% of the original C-C bond length of 1.53 Å. The inset in the upper left shows the polyacrylonitrile chain after failure using a cutoff distance $Rmax \geq 2.6$ Å. Cutoffs below 2.6 Å cause premature failure, particularly for a short cutoff at 1.7 Å (arrows in red). (b) Stress-strain curve for $R_{max}$ values from 2.6 Å to 4.0 Å, equal to 170% to 261% of the original C-C bond length of 1.53 Å. All bond cutoffs in this range show almost identical failure and correspond to equilibrium conditions in the IFF-R simulation. (c) Total energy vs strain curves

for bond cutoffs from 1.7 Å to 2.6 Å shown in (a). The total energy spikes instantaneously for a $R_{max}$ of 1.7 Å and shows discontinuities for $R_{max}$ values up to 2.6 Å, near the respective strain at break. (d) Total energy vs strain curves for bond cutoffs from 2.6 Å to 4.0 Å shown in (b). The total energy curves are identical, indicating equilibrium bond dissociation. Small, similar discontinuities at break are seen. Overall, a minimum bond cutoff of at least 170% of the equilibrium bond distance of C-C single bonds is recommended.

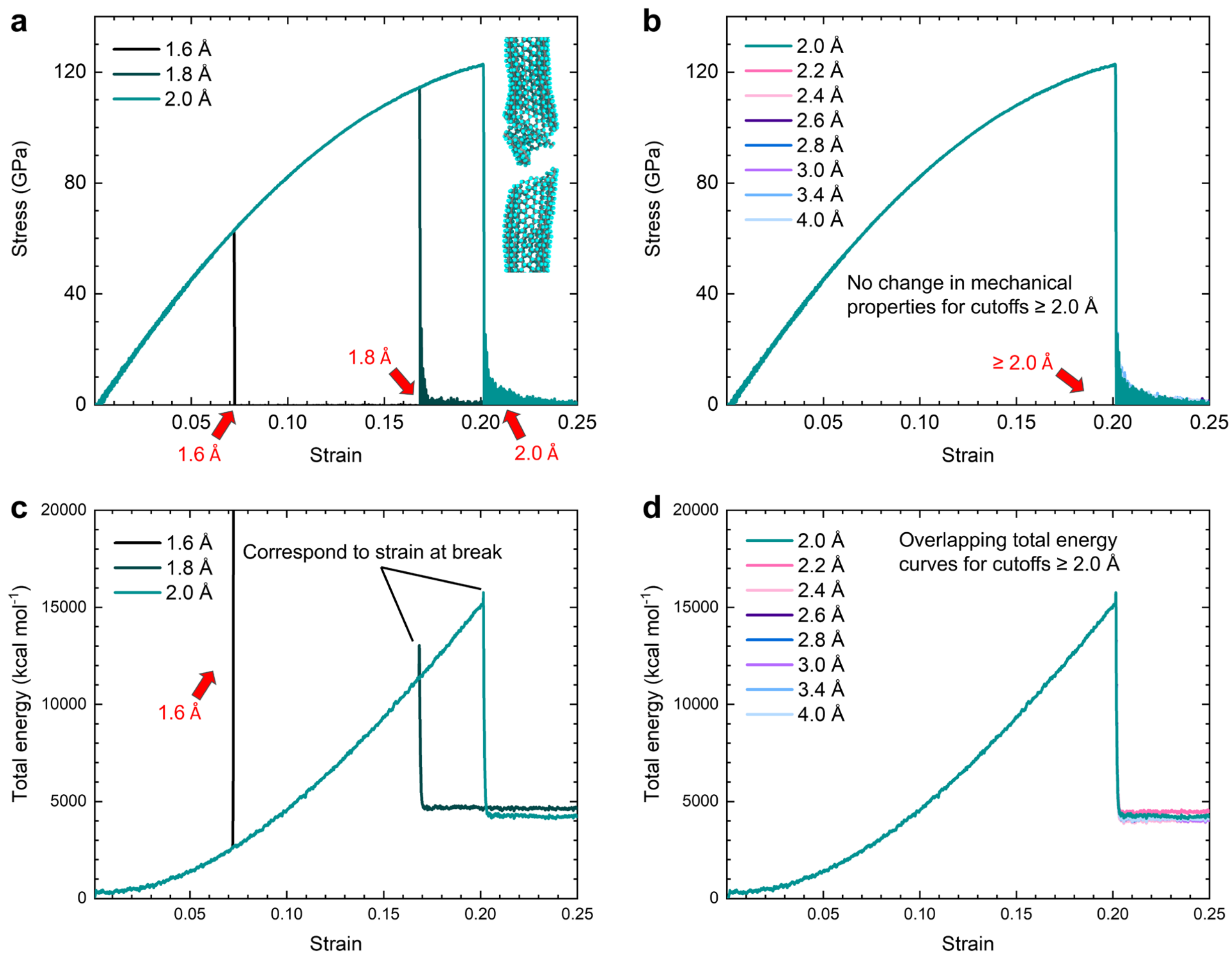

**Figure S4.** Stress-strain curves and the total energy of a single wall carbon nanotube (SWCNT) for 10 different bond cutoffs ($R_{max}$) using a shifted Morse potential (by $-D_{ij}$). The bond/break LAMMPS command was used during a tensile simulation. (a) Stress-strain curve for $R_{max}$ values from of 1.6 Å to 2.0 Å, equal to 113% to 141% of the original $C_{ar} - C_{ar}$ bond length of 1.42 Å. Cutoffs below 2.0 Å cause premature failure (arrows in red). The inset in the upper left shows a SWCNT after failure with cutoff $Rmax \geq 2.0$ Å. (b) Stress-strain curve for $R_{max}$ values from 2.0 Å to 4.0 Å, equal to 141% to 282% of the original C-C bond length of 1.42 Å. All bond cutoffs in this range show almost identical failure and correspond to equilibrium conditions in the IFF-R simulation. (c) Total energy vs strain curves for bond cutoffs from 1.6 Å to 2.0 Å shown in (a). The total energy shows discontinuities near the respective strain at break for the different $R_{max}$

values. (d) Total energy vs strain curves for bond cutoffs from 2.0 Å to 4.0 Å shown in (b). The total energy curves are nearly identical, indicating equilibrium bond dissociation. Small, similar discontinuities at break are seen. Overall, a minimum bond cutoff of at least 140% of the equilibrium bond distance of $C_{ar}$–$C_{ar}$ aromatic bonds (1.53 Å) is recommended.

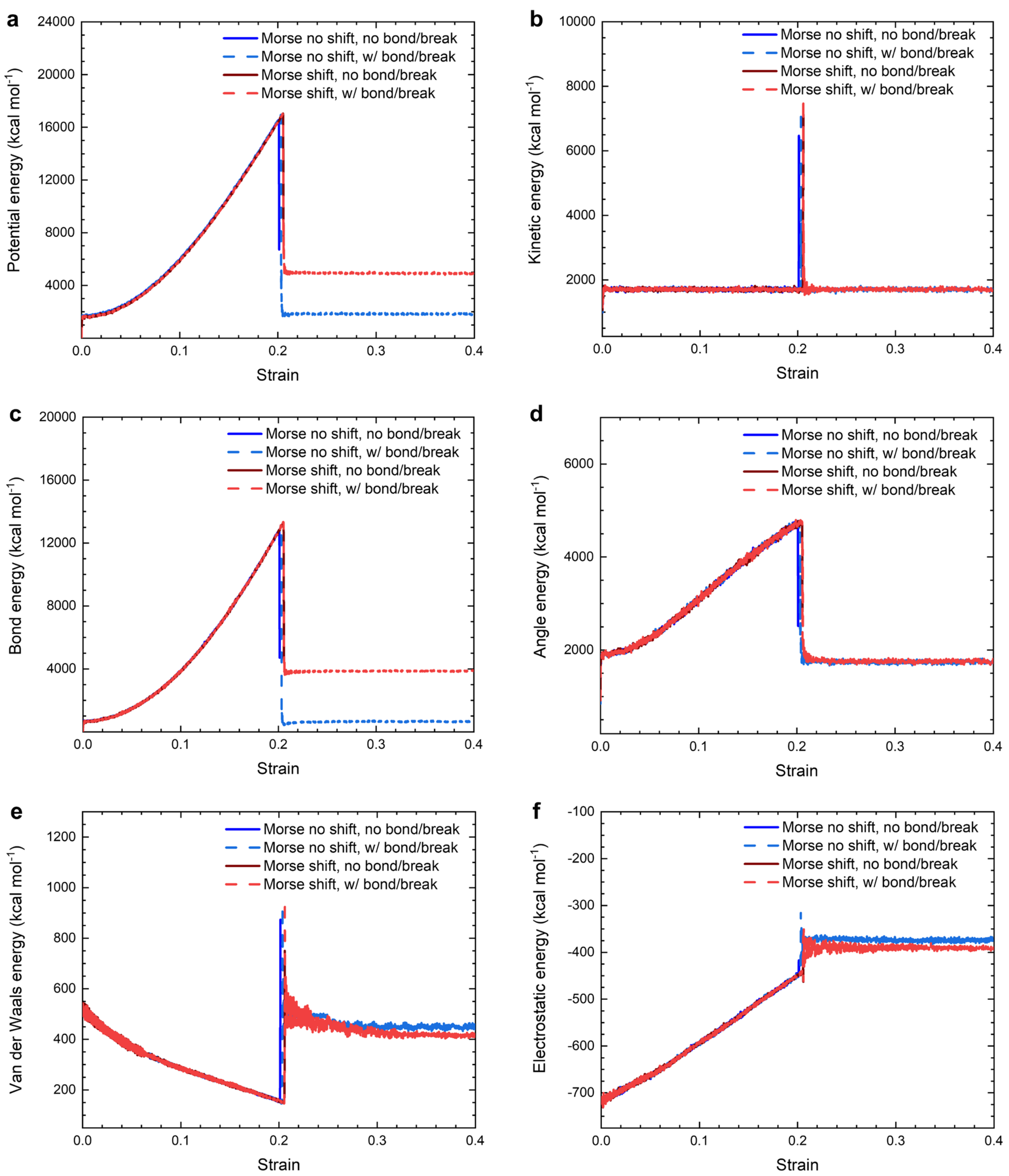

**Figure S5.** Energy contributions as a function of strain during simulations of tensile stress-strain curves of a SWCNT using IFF-R Morse bond parameters. The potential energy and bond energy for the default Morse potential and the shifted Morse potential have been normalized to 0 kcal/mol

for comparison. Differences between the default Morse potential, the shifted Morse potential (by $D_{ij}$), and the default and shifted Morse potential with bond break ($Rmax_{Car\text{-}Car}$ = 2.84 Å) can be seen. Contributions from the torsion potential (that stabilizes a flat $sp^2$ geometry) are not shown. One of the future challenges is the reduction, or elimination, of discontinuities in energy upon bond breaking without adding artificial new parameter space. (a) The total potential energy, (b) the kinetic energy, (c) the bond energy, (d) the angle energy, (e) the van der Waals energy, and (f) the electrostatic energy. The bond energy (a, c) continues to show discontinuities upon dissociation since also neighbor bonds are strained and released in the process of disconnecting the strained Morse bond, as well as angles, torsions, and nonbond interactions in the vicinity of the reaction site (d, e, f). The kinetic energy temporarily increases due to gained mobility (b). The Coulomb energy increases due to repulsion between like charges after bond C-C scission (f). The effects depend on whether new atom types and force field parameters are assigned after bond scission.

| Bond type | Dissociation mechanisms | Parameterization recommendations |
|---|---|---|
| Graphitic | 1 | Update nonbonded parameters for dissociated carbon atoms |
| Cellulose | 2a | Update charge distribution for carbon and oxygen atoms |
| | 2b | Create new chemical bonds with two C atoms redefined as sp$^2$/alkene |
| | 2c | Create new chemical bond and redefine R-O$^-$ oxygen to hydroxyl oxygen atoms |
| PAN | 3a | Update nonbonded parameters for dissociated carbon atoms |
| | 3b | Update charge distributions for dissociated carbon atoms |

**Figure S6.** Bond dissociation mechanisms for $C_{ar}$-$C_{ar}$ graphitic bonds, C-O bonds in cellulose, C-C bonds in polyacrylonitrile and suggestions to parameterize the newly dissociated atoms (ref. [1]). Proposed cleaved bonds are shown in red and R designates substituent groups not participating in the reactions/dissociation. Various other follow-on reactions of radicals and ions are possible, e.g., with oxygen, water from moisture, and atmospheric gases depending on the reaction conditions. Mechanism 1 indicates homolytic cleavage of $C_{ar}$-$C_{ar}$ bonds in graphitic structures and carbon nanotubes into radicals. Mechanism 2a shows the heterolytic cleavage of C-O ether linkages in cellulose. If bond formation can be considered in the simulation, mechanism 2b proposes the stabilization of carbocation intermediates through the formation of a double bond, and mechanism 2c the protonation of alcoholate groups. Mechanism 3a proposes homolytic dissociation of C-C bonds in PAN, and mechanism 3b heterolytic dissociation, both of which could lead to multiple

products. All proposed follow-on reactions and parameterization recommendations can be handled via the LAMMPS REACTER package (ref. [2]).

**a** DETDA resonance

| Bond | No. of single bonds | No. of double bonds | Double bond character (%) |
|---|---|---|---|
| 1, 2 | 7 | 3 | 30 |
| 2, 3 | 7 | 3 | 30 |
| 3, 4 | 8 | 2 | 20 |
| 4, 5 | 5 | 5 | 50 |
| 5, 6 | 5 | 5 | 50 |
| 6, 1 | 8 | 2 | 20 |

**b** 4,4'-DDS resonance

| Bond | No. of single bonds | No. of double bonds | Double bond character (%) |
|---|---|---|---|
| 1, 2 | 5 | 2 | 28.6 |
| 2, 3 | 2 | 5 | 71.4 |
| 3, 4 | 6 | 1 | 14.3 |
| 4, 5 | 6 | 1 | 14.3 |
| 5, 6 | 2 | 5 | 71.4 |
| 6, 1 | 5 | 2 | 28.6 |

**Figure S7.** Resonance structures of (a) DETDA and (b) 4,4'-DDS with atom labels (top) and estimated double bond character of individual bonds. Higher double bond character correlates with increased bond dissociation energy. The percentage of double bond character for each bond was calculated as the fraction of resonance structures showing double bonds. In DETDA, double bond character increases in the order 3,4 < 1,2 < 5,6. In 4,4'-DDS, the double character increases in the order 3,4 < 1,2 < 2,3 with high double bond character of the 2,3 and 5,6 bonds (>70%) due to a push-pull effect from the p-situated amino and sulfoxide groups.

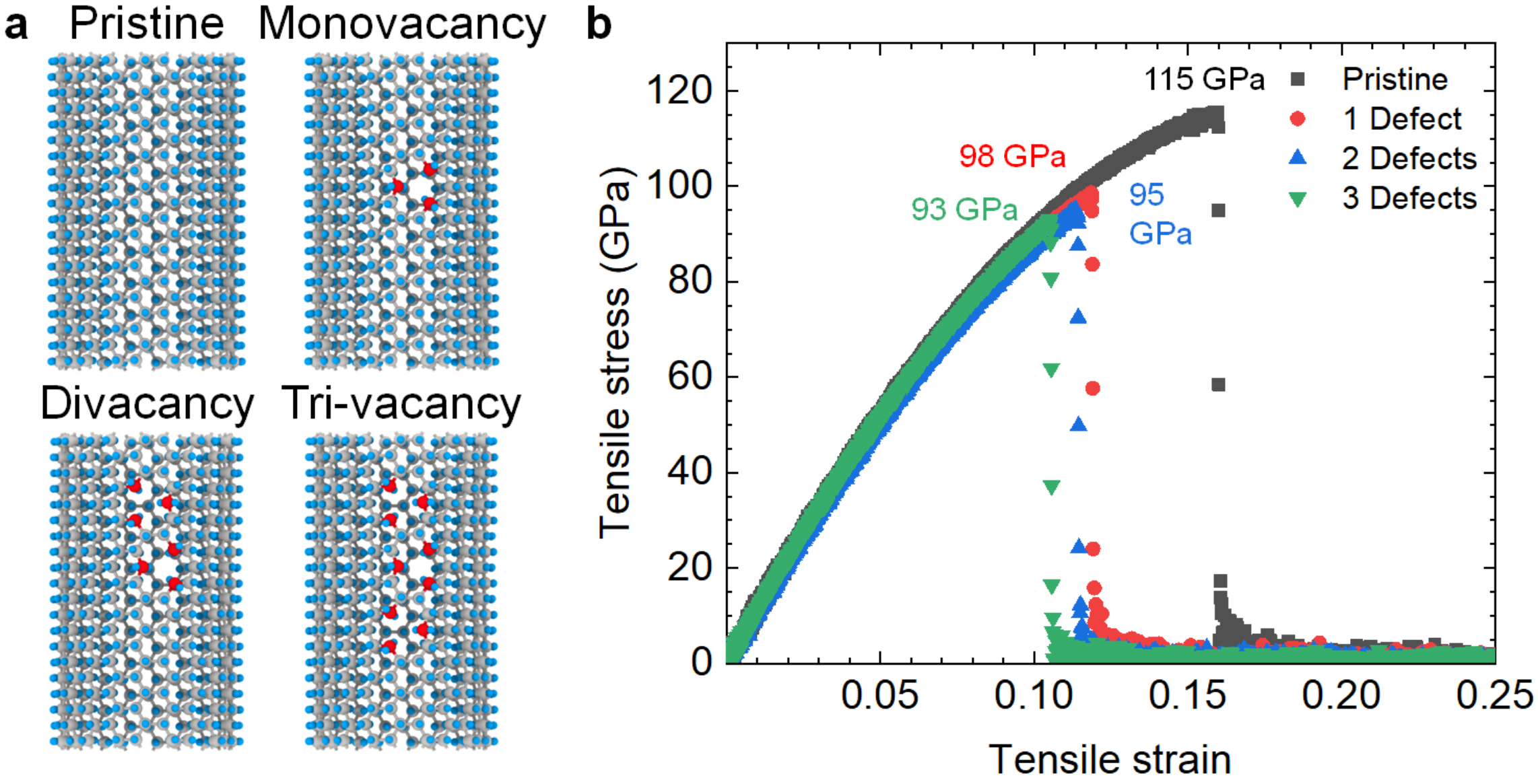

**Figure S8.** Demonstration of the variability of the stress-strain curves and failure mechanisms of single-wall carbon nanotubes containing defects as seen in MD simulations with IFF-R. (a) Models of 2.0 nm diameter (15,15) SWCNTs with increasing number of missing atoms from 0 (pristine) to 3 (tri-vacancy). Sites of missing atoms are highlighted in red. (b) Stress-strain curves for the carbon nanotubes in (a). A limit for the influence of missing atoms on the reduction of tensile strength and ultimate strain can be seen. A single missing atom from a pristine SWCNT structure causes the greatest reduction in tensile strength of -14.7%, i.e., from 115 GPa to 98 GPa. Subsequent removal of atoms leads to reductions of -3.1% and -2.1%, respectively.

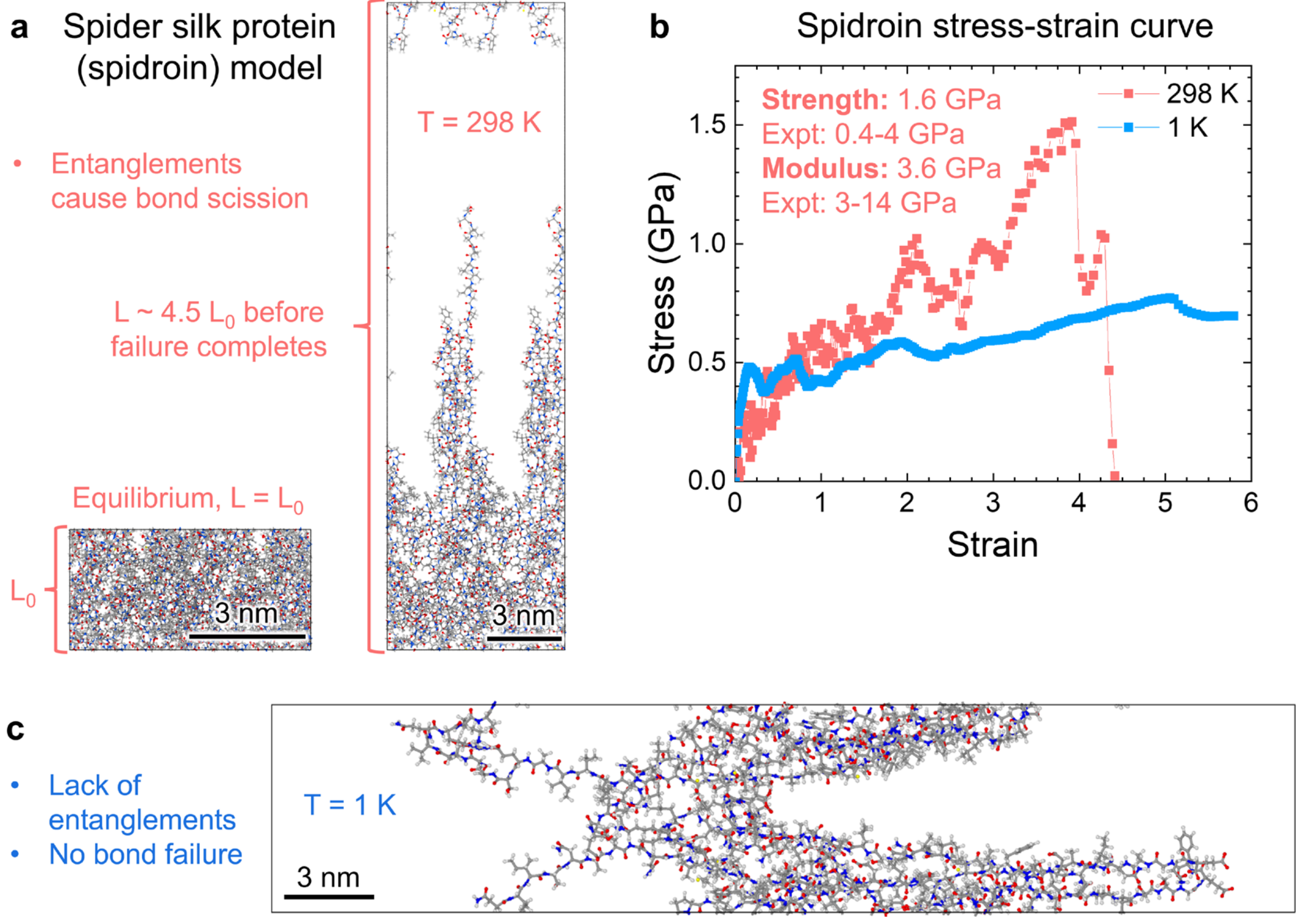

**Figure S9.** Use of CHARMM36m parameters[3] with IFF-R Morse bonds for the simulation of tensile failure of the spider silk protein spidroin. (a) A model of spidroin consisting of two protein strands in equilibrium (left) and at an elongation of more than 4 times the original box length (right) after failure at a temperature of 298 K. The dynamics at 298 K led to the formation of entanglements of the chains, which affect the failure mechanism. (b) Stress-strain curve of the spidroin model obtained by reactive molecular dynamics simulations. In the chosen model configuration, spidroin showed a tensile strength of 1.6 GPa and a Young's modulus of 3.6 GPa. The values are in the range of the experimentally determined tensile strength of 0.4 to 4.0 GPa and the experimentally measured tensile modulus of 3 to 14 GPa for spider silk (refs. [4-7]). No modifications of CHARMM36m parameters[8] for spidroin, other than the introduction of IFF-R Morse bonds were made to obtain the data reported. (c) The response of spidroin to tensile stress at a lower temperature of 1 K. Entanglements of the protein strands were not observed due to the limited mobility of the amino acids at 1 K and led to different mechanical properties. Using the same starting configuration as at 298 K, the two spidroin molecules separated during the tensile simulation, the stress response was dominated by nonbonded interactions, and bond breaking did not occur. At either temperature, the mechanical properties are also influenced by the nanoscale morphology, relaxation time, and strain rate, which were not further examined here.

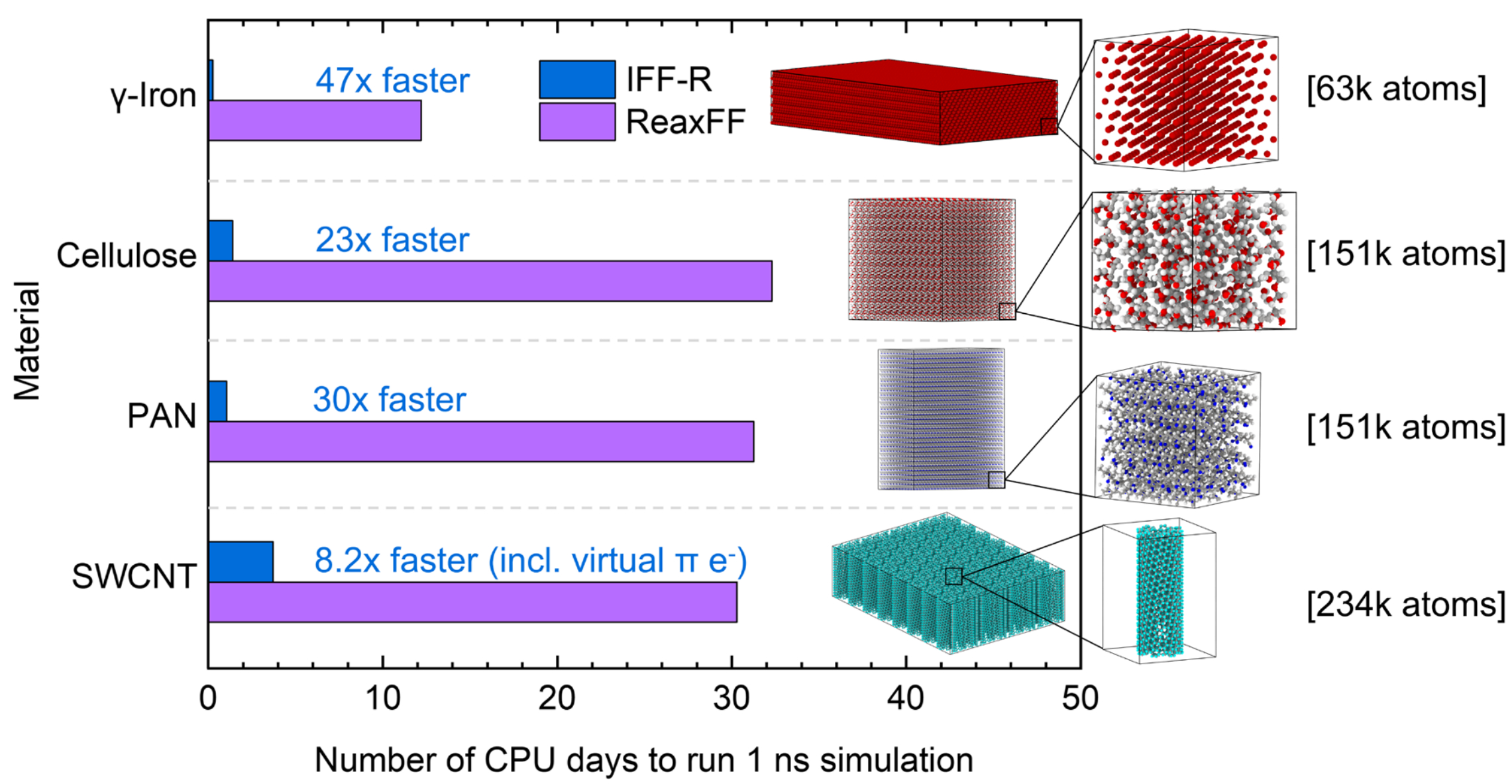

**Figure S10.** Comparison of CPU time needed for the simulation of stress-strain curves using IFF-R and ReaxFF for large model sizes of 63,000 to 234,000 atoms. CPU time and snapshots at failure are shown for a crystal of γ-iron, a crystal of cellulose Iβ (ref. [9]), a crystal of syndiotactic poly(acrylonitrile) (PAN), and bundle of single-walled carbon nanotubes (SWCNT) of 1 nm diameter using IFF-R (PCFF-R in case of cellulose). The materials are the same as in Figure 4, however, scaled 50-100x larger for γ-iron and the polymers, and ~20x larger for the CNT bundle. Full scale images of the model systems are shown in the chart area and close-up images on the right-hand side of the graph. The benchmark simulations were carried out on 24 CPUs, using 1000 steps of molecular dynamics simulations with a constant strain rate in the tensile direction, and extrapolated for 1 ns duration. The NPT ensemble was employed with pressure control in $x$ and $y$ directions, and no pressure control in the tensile ($z$) direction. IFF-R completes 1 ns molecular dynamics simulation using 24 CPUs in 6 hours to 4 days for all 4 materials. The ReaxFF simulations would finish between 12 days and 1 month. The efficiency of IFF-R is about the same as for small model systems, including a slight decrease for cellulose Iβ and γ-iron crystals, and a slight increase for PAN and CNTs.

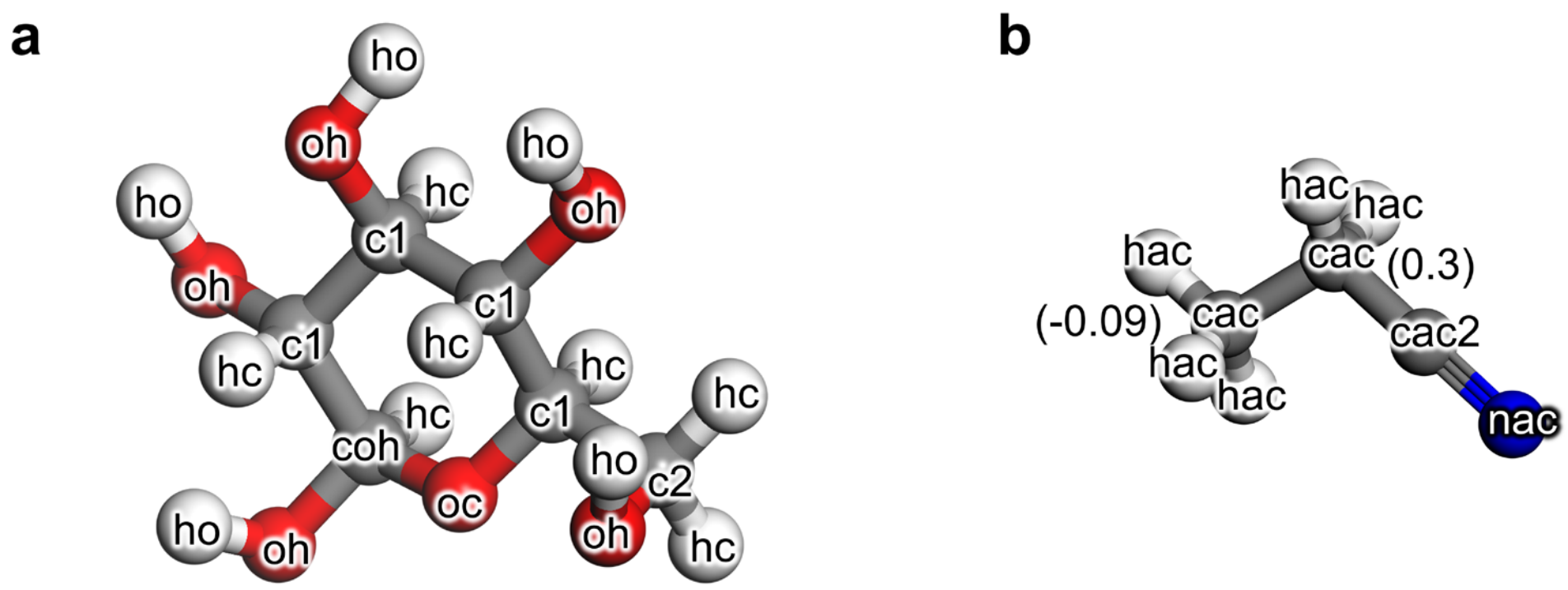

c

| Atom Type | 9-6 Lennard-Jones potential | | Atomic charge (e) |
|---|---|---|---|
| | $r_{min}$ (Å) | Epsilon (kcal mol$^{-1}$) | |
| **Glucose** | | | |
| ho | 1.098 | 0.013 | 0.45 |
| hc | 2.7 | 0.02 | 0.03 |
| oh | 3.4 | 0.28 | -0.7 |
| oc | 3.4 | 0.18 | -0.5 |
| coh | 4.0 | 0.05 | 0.47 |
| c1 | 4.0 | 0.06 | 0.22 |
| c2 | 4.0 | 0.06 | 0.19 |
| **Propionitrile** | | | |
| hac | 2.8 | 0.012 | 0.03 |
| cac | 4.0 | 0.06 | -0.09/0.30 |
| cac2 | 3.8 | 0.05 | 0.45 |
| nac | 4.2 | 0.24 | -0.79 |

**Figure S11.** Models, force field types, and nonbonded parameters for α-D-glucose and propionitrile. (a, b) The structure of α-D-glucose and propionitrile with atom types (force field types). The model of (b) propionitrile also lists the charges on the carbon backbone. (c) Table of force field types, 9-6 Lennard-Jones parameters, and atomic charges consistent with dipole/multipole moments. The parameters for α-D-glucose are PCFF-R and include updated atomic charges. The parameters for propionitrile are IFF-R (higher quality).

**Table S1.** Morse parameters for selected bonds in IFF-R and the expected range for any chemistry (bottom two lines). Origin of reference data: Experimental data for equilibrium bond lengths are available in ref. [10] and for bond dissociation energies in refs. [10-13]. Data for all types of carbon-carbon bonds additionally consider ref. [14], and the data for carbon-carbon bonds in graphite and carbon nanotubes also consider ref. [15]. The range of bond energies across a broad range of chemistries from refs. [10-14] can be significantly narrowed down once the specific chemical environment of the bonds is considered (<3% uncertainty).

| **Bond type** | **Equilibrium bond length $r_{0, ij}$ (Å)** | | **Width $\alpha$ (Å$^{-1}$)[b]** | **Bond dissociation energy $D_{ij}$ (kcal/mol)** | |
|---|---|---|---|---|---|
| | **IFF-R** | **Experiment[10]** | | **IFF-R** | **Experiment[10-13]** |
| $C_{ar}$ – S (sulfone) | 1.80 | 1.77 | 1.9 | 75 | 75-85 |
| $C_{ar}$ – N (amine) | 1.39 | 1.38 | 2.0 | 110 | 100-110 |
| C – C (single bond) | 1.53 | 1.54 | 2.2 | 85 | 70-90 |
| $C_{ar}$ – $C_{ar}$ (benzene) | 1.39 | 1.38 | 2.2 | 150 | ~155 |
| $C_{ar}$ – $C_{ar}$ (graphitic and CNTs) | 1.42 | 1.42 | 2.4 | 124 | 108-140 |
| C – O (ether)[a] | 1.42 | 1.43 | 2.0 | 85 | 81-98 |
| C = C (double bond)[a] | 1.34 | 1.34 | 1.9 | 152 | 140-172 |
| Typical range | 0.7-2.8 | 0.7-2.8 | 1.7-2.4 | 0-250 | 0-250 |
| Typical uncertainty | <1% | <1% | <0.1 | <10% | <10% |

[a] The Morse bond parameters for C=C double bonds and C-O ether bonds are generic and require fine-tuning according to specific chemical environments (see examples in Figure 2). [b] The alpha value may vary due to substituent effects not specifically included here.

**Table S2.** Comparison of surface and vaporization energies for graphite, propionitrile, and α-D-glucose using experimental data, computational data from IFF-R/PCFF-R, and two different ReaxFF parameter sets by Damirchi et al. (ref. [16]) and Liu et al. (ref. [17]).

| Material | Expt | IFF-R | Dev (%) | ReaxFF (Damirchi) | Dev (%) | ReaxFF (Liu) | Dev (%) |
|---|---|---|---|---|---|---|---|
| **Surface energy ($mJ/m^2$)** | | | | | | | |
| Graphite | 186 ±2.0[18,19] | 188 ± 1.9 | +1.1 | 207 ± 1.9 | +11 | 242 ± 2.0 | +30 |
| **Vaporization energy of monomer (kJ/mol)** | | | | | | | |
| Propionitrile | 36.1 ±1.6[13] | 37.0 ± 0.35 | +2.5 | 29 ± 0.8 | -20 | 47 ± 0.9 | +30 |
| α-D-glucose | 181.7 ±5.0[20] | 176 ± 0.33 | -3.1 | 118 ± 2.5 | -35 | 169 ± 1.6 | -7.0 |

## S1. Detailed Computational Methods

**S1.1. Scans of Bond Dissociation Energies.** *Model Building.* The amine molecules were aligned in the *x*-*y* plane of the simulation cell and the bonds of interest aligned along the *x*-axis to easily modify the interatomic distance. The bond length between atoms *i* and *j* was elongated with a step-size of 0.1 Å or less while constraining the positions of all other atoms.

*DFT (Density Functional Theory) and MP2 Protocol.* DFT calculations use the B3LYP gradient-corrected hybrid functional,[21,22] and MP2 calculations were based on Hartree-Fock theory with added electron correlations up to second order perturbation theory. MP2 differs from DFT in including corrections to the wavefunction, providing more accurate bond scans than DFT to derive IFF-R parameters. The Gaussian 2016 program[23] was used and employed an all-electron Gaussian basis set to describe electronic configurations of molecules. For the constituent atoms of amine monomers, the standard split-valence polarized double-zeta basis sets 6-31G (d, p) were used. At each point of the energy surface, the convergence criteria for the RMS density matrix and total energy were $10^{-8}$ and $10^{-6}$ eV, respectively. No periodic boundary conditions were applied to calculate bond dissociation energies of the molecules. For simplicity, the energy of the equilibrium molecular configuration was set to zero.

*Molecular Mechanics Protocol.* Simulations of bond dissociation curves using IFF-R, as well as the ReaxFF parameter set by Singh et al.[24] were carried out by a series of energy minimizations in LAMMPS using the Polak-Ribiere version of the conjugate gradient algorithm.[25] The minimization stopping criteria were set to a $10^{-4}$ energy tolerance, or $10^{-6}$ kcal/(mol·Å) force tolerance, within a maximum of 10,000 steps. During energy minimization at each bond distance, the positions of all atoms were constrained except for the two atoms *i* and *j* involved in bond dissociation. The equilibrium energies of the molecule were recorded as a function of bond length and normalized to zero at the distance of lowest energy.

**S1.2. Simulation of Tensile Modulus and Strength.** *Model Building and Parameter Choices.* A model of an (8, 8) armchair single-wall carbon nanotube with a diameter of 1.085 nm and a length of 4.92 nm (i.e., 20 repeat units in length) was created in Materials Studio. Virtual π-electrons were added for stress-strain simulations with IFF and IFF-R,[26] and removed for simulations with ReaxFF (which does not explicitly represent such features).

A model of crystalline, syndiotactic PAN was created using the "Build Homopolymer" tool in Material Studio. The model matched the experimentally known density[27] of 1.16 $g/cm^3$ and assumed a rectangular crystalline supercell with parallel chains and lattice dimensions of 3.12 nm x 3.07 nm x 3.42 nm. No experimental information on unit cell parameters and atomic positions, e.g., from X-ray diffraction, was available for syndiotactic 100% crystalline PAN.

A model of crystalline cellulose Iβ was created using the structure built by Nishiyama et al. based on X-ray and neutron diffraction data.[9]

A unit cell for γ-iron was built using the IFF model database with initial estimated density of 8.18 $g/cm^3$ and expanded into a supercell of the dimensions 1.77 nm x 2.18 nm x 1.85 nm.[28] γ-iron is a suitable model for stainless steel, typically including minor components of other alloying elements (Cr, Ni, C, Mo, Mg). Otherwise, pure iron assumes a bcc structure of α-iron at room temperature.[29] The force field for simulations was chosen as IFF and IFF-R with 9-6 Lennard-Jones options.[30,31]

We further carried out simulations with ReaxFF for comparison using the same models. Stress-strain simulations of the SWCNT, PAN crystal, and the cellulose Iβ crystal utilized the ReaxFF parameters developed by Liu et al.[17] and Damirchi et al.[16] For iron, the ReaxFF parameters by Islam et al.[32] were used, which were originally developed for analyzing hydrogen embrittlement in pure and defective iron.

*Simulation Protocol.* Stress-strain simulations under a tensile load were performed by assigning Morse parameters to the bonds that were considered the weakest according to their bond dissociation energies, for example, C-S bonds rather than C=C (double) bonds. The Morse parameters for $C_{ar}$-$C_{ar}$ partial double bonds were used for graphite and CNTs, the Morse parameters for C-C single bonds were used for the PAN crystal, and Morse bond parameters for cellulose included C-C single-bonds and C-O single bonds. No Morse bond parameters were needed for iron since it only involves Lennard-Jones potential and no bonded terms in IFF. The inclusion of a metal remains important here to showcase how IFF-R is applicable to all types of materials and their mixtures, even when no Morse bonds are required due to the specific type of bonding.

The simulation of tensile stress-strain curves involved equilibration of the model systems using molecular dynamics simulations in the NVT ensemble at 298 K for 10 picoseconds. Following equilibration, molecular dynamics simulations were carried out in the NPT ensemble to obtain the stress-strain curves, except for the SWCNT which was run using the NVT ensemble due to inclusion of some vacuum around the CNT (the cross-sectional area pertaining to the SWCNT only was subsequently calculated). Simulations of the stress-strain curves under tensile load employed 1 atm pressure in the lateral directions and a constant rate of strain of $2 \cdot 10^{-6}$ engineering strain every 10 timesteps of 0.5 fs along the tensile direction (0.4 per ns). Some simulations used higher strain rates of up to 1.0 $ns^{-1}$ which did not change the results. γ-iron was strained in the [111] direction perpendicular to the (111) surface. The molecular dynamics simulations involved a timestep of 0.5 femtoseconds to allow small, equilibrium displacements and accurate recordings of strain, stress, and energy. The simulations were continued until material failure, or a maximum of 100% strain.

*Analysis.* The stress was recorded as a function of strain to draw the computed stress-strain curves. The stress was obtained according to the virial theorem relative to the initial cross-sectional area of the box perpendicular to the tensile direction for all systems except the SWCNT. The model system of the SWCNT included vacuum in the simulation box, and the stress relative to the cross-sectional area of the simulation box was normalized to the cross-sectional area of the CNT only, excluding the contribution by vacuum in the simulation cell. The normalized area (in nm) was obtained using the equation $A = N\pi(r + 0.19\,nm)^2$ where $N$ is the number of CNTs, $r$ is the initial (equilibrium) radius of a CNT, and 0.19 nm corresponds to the addition of the van-der-Waals radius (equal to half of the layer spacing of graphite of 0.17 nm, plus a small allowance of 0.02 nm to account for the non-planar geometry and hexagonal packing of CNTs). The engineering strain was calculated as the relative expansion of the simulation cell in the tensile direction relative to the original length.

The Young's modulus was calculated by applying a line of best fit to the linear stress-strain response of the materials models near zero stress, typically between 0.00 to 0.01 tensile strain, and the slope of the best fit line was determined to be the Young's modulus ($E = \Delta\sigma/\Delta\varepsilon$). The tensile strength was defined as the maximum stress at or before material failure.

The MD simulation trajectories were visualized using the OVITO v2.8 software.[33] Inspection of all trajectories as a function of simulation time ensured the absence of unexpected flaws in the models and associated stress-strain curves for analysis (e.g. human error through missing atoms or in setup protocol). Simulations were repeated in case issues were found. Atomic coordinates were printed to the trajectory file every 1000 timesteps. To visualize the exact instance of bond failure, a separate simulation was carried out in which the models were strained up to approximately 2000 timesteps before failure. Afterwards, atomic coordinates were printed every 1 to 10 timesteps to capture the instance of initial bond dissociation.

**S1.3. Ab-Initio Calculation of the Young's Modulus for Polyacrylonitrile.** We utilized the CASTEP[34] program to compute the Young's modulus of syndiotactic, crystalline PAN in the direction of alignment of the chains as a reference relative to MD simulations with IFF-R. First, a geometry optimization was applied to an orthorhombic simulation cell with dimensions 1.00 nm x 1.15 nm x 1.16 nm and 4 aligned syndiotatic PAN chains with a density of 1.16 g/cm$^3$.[27] The energy cutoff for the geometry optimization was set to 280 eV with a self-consistent field tolerance of 1.0e-6 eV/atom. The fast Fourier transform grid was set to a density of 50x72x96 with a scaling factor of 1.2 and a 1x2x1 k-point grid was utilized.

The CASTEP analysis tool in Materials Studio was then used to calculate the elastic constants of the optimized PAN structure using the GGA PBE density functional and fine quality results using a 2x2x2 Monkhorst-Pack grid of k-points. The simulation involved 4 steps for each strain with a maximum strain amplitude of 0.003. The lattice parameters were allowed to change, and convergence thresholds included an energy tolerance of 2.0e-6 eV/atom with maximum force of 0.006 eV/Å and a maximum lattice displacement of 2.0e-4 Å. Simulations were run until convergence after 120 iterations.

**S1.4. Simulation of the Surface Energy of Graphite and γ-Iron.** *Model Building.* An initial model of bulk graphite with 3D periodicity was created in Materials Studio with lattice parameters $a$ = 1.476 nm, $b$ = 1.2874 nm, $c$ = 4.08 nm, and $\alpha = \beta = \gamma = 90°$ (a rectangular supercell corresponding to the hexagonal unit cell). The equilibrium density was calculated to be 2.24 g/cm$^3$, and the spacing between graphitic layers was 0.335 nm.[18,19,35] A model of bulk γ-iron was created using the lattice parameters $a$ = 1.76505 nm, $b$ = 2.1837 nm, $c$ = 1.85295 nm, and $\alpha = \beta = \gamma = 90°$, corresponding to a supercell with the $z$-axis aligned in the [111] direction and a density of 8.18 g/cm$^3$ at room temperature as given in ref. [28].

Cleaved surface slabs of graphite and γ-iron were created after obtaining equilibrium lattice parameters for each material from an NPT ensemble simulation at 298 K and 1 atm pressure for 1 ns. The equilibrated structures were imported into Materials Studio, and the simulation cell was expanded by 4.0 nm in the z-direction, creating a solid-vacuum interface. As a result, two surfaces of equal dimension were created by introducing a vacuum slab into the 3D periodic model.

*Simulation Protocol.* First, the models were subjected to energy minimization with an energy tolerance of $10^{-4}$ and a force tolerance of $10^{-6}$ kcal/(mol·Å). Second, molecular dynamics simulations at 298 K and 1 atm pressure in the NPT ensemble were performed for 1 ns to obtain structures in equilibrium and average lattice parameters. Third, MD simulations were continued in the NVT ensemble at 298 K for 1 ns using the average lattice parameters and utilized to obtain the total average energy of the equilibrated bulk material and for the cleaved structure. The LJ cutoff was set to 12.0 Å (IFF standard) and electrostatic interactions were computed in high accuracy using the PPPM kspace solver with an accuracy of $10^{-4}$ (and including a cutoff of 8.0 Å or higher for the direct summation, which has no influence on the results).

*Analysis.* The time-averaged total energy for the last portion of 1 ns in equilibrium was recorded as $E_{bulk}$ and $E_{cleaved}$, and then used to calculate the surface energy of graphite and γ-iron, whereby $A$ was the surface area of the slabs

$$\gamma = \frac{E_{cleaved} - E_{bulk}}{2\ x\ A}. \quad \text{(S1)}$$

The surface energies are reported in mJ/m$^2$.

**S1.5. Simulation of the Enthalpy of Vaporization of Propionitrile and the Enthalpy of Sublimation of α-D-Glucose.** *Model Building.* The dimensions of a model of liquid propionitrile were determined using an orthogonal simulation box containing 108 $CH_3$-CN molecules, leading to equilibrium dimensions of $a$ = 2.328 nm, $b$ = 2.328 nm, and $c$ = 2.330 nm. These dimensions reproduced propionitrile's experimentally determined density of 0.782 g/cm$^3$ within 0.5%.[10] Simulations of the vapor phase assumed a large orthogonal simulation cell with dimensions of $a$ = $b$ = $c$ = 63.0 nm containing 20 propionitrile molecules.

A model of the crystal structure of α-D-glucose was created using a superstructure based on the unit cell dimensions measured by McDonald et al.[36] The supercell of crystalline α-D-glucose contained 64 glucose molecules and had lattice parameters of $a$ = 1.858 nm, $b$ = 2.530 nm, $c$ = 2.680 nm. The simulation cell for α-D-glucose in the vapor phase contained 20 glucose molecules and had lattice parameters of $a$ = $b$ = $c$ = 100.0 nm.

*Simulation Protocol and Analysis.* MD simulations of liquid propionitrile and crystalline glucose were carried out in the NPT ensemble for 1 ns (following longer initial equilibration). Simulations of the vapor phases were carried out for 1 ns in the NVT ensemble. The enthalpy of vaporization or sublimation enthalpy $\Delta H_{v,s}$, respectively, was calculated as follows:

$$\Delta H_{v,s} = \frac{\langle E_{vap} \rangle}{\#\ of\ vap\ mol.} - \frac{\langle E_{liq,cry} \rangle}{\#\ of\ liq, cry\ mol.} + RT \qquad \text{(S2)}$$

$E_{vap}$ was the average total energy of the simulation system in the vapor phase, $E_{liq,cry}$ was the average total energy of the liquid or crystal simulation system, and $RT$ is the gas constant 8.314 J/mol multiplied by the simulation temperature. The formula includes the necessary correction of the molar vaporization energy to the molar vaporization enthalpy ($\Delta H = \Delta U + PV$), using the ideal gas law $pV = nRT$ to approximate the volume work under constant pressure (n = 1 mol).[37, 38]

**S1.6. Parameterization of the Morse Bond Potential for Carbon Nanotubes and Polyacrylonitrile.** The Morse bond parameter α of graphite and SWCNTs for IFF-R was iteratively obtained by computing the vibration spectra (including the Raman G band), comparison to experimental data, and adjustments of α for a close match (<20 $cm^{-1}$ deviation). The refinements of α also improved the predicted Young's Modulus and ultimate tensile strength relative to experimental data[39] and quantum calculations.[34, 39-42] The final value of α = 2.4 $Å^{-1}$ was chosen to reproduce the mechanical properties and the vibration spectra well.

Similarly, small adjustments of α helped to closely match the Young's modulus and tensile strength of PAN in the simulations. However, the correlation of α rests with the reproduction of IR and Raman spectra as mechanical properties are additionally influenced by the bond stiffness (α), other bonded, as well as nonbonded forces.

**S1.7. PCFF Parameters for Propionitrile and PCFF Parameter Improvements for Glucose.** Improvements of the base parameters in PCFF (not Morse parameters) were made for propionitrile and α-D-glucose (Figure S11). The atomic charges and the nonbonded parameters of propionitrile were reevaluated using consistent dipole moments of small nitrile molecules[10] and a series of molecular dynamics simulations to adjust LJ parameters using the Discover program in Materials Studio.[43] The overall performance of the updated PCFF parameters for propionitrile is close to IFF standards. The base parameters for glucose in PCFF seriously lack structural and thermodynamic consistency and were updated for partial improvement, still far from IFF standards, since the development of basic force field parameters is not the focus of this work.

Model building of propionitrile and α-D-glucose followed the protocols described in Section S1.5. The simulation of liquid and vapor phases of propionitrile was carried out using 3D boxes with 108 molecules at a temperature of 298.15 K, velocity scaling for temperature control, a time step of 0.5 fs, and a total simulation time of at least 1 ns. The simulations utilized the NPT ensemble for the liquid phase, maintaining constant pressure at 1.00 atm controlled by the Parrinello-Raman barostat. The NVT ensemble was used for the gas phase, including large distances between individual molecules of >2 nm. The summation of Lennard-Jones interactions was carried out with a spherical cutoff at 12.0 Å and electrostatic interactions were computed as an Ewald summation with a high accuracy of $10^{-4}$.

For analysis, the average density, the average total energy for liquid propionitrile and the average energy for propionitrile in the gas phase were calculated. The enthalpy of vaporization for propionitrile was obtained using equation (S2). Iterative modifications of the atomic charge distribution of proprionitrile and of the 9-6 Lennard Jones parameters were made until the liquid density and enthalpy of vaporization were within the error of experimental observations (Table 3, Figure S11, and Table S2).[10,13] For example, the atomic charge on the C atom "cac" next to the nitrile group had a significant effect on the vaporization energy. The default atomic charge was +0.11 e while, to reproduce the liquid density and vaporization enthalpy known from measurements, a charge of +0.30 e, which is about 3 times the default charge, was necessary (Figure S11). Likewise, $r_{min}$ and epsilon parameters in the Lennard-Jones potential were adjusted to account for changes in atomic radii and polarizability consistent with similar chemistry covered in IFF[44] and other force fields (CHARMM, OPLS-AA).

Improvements of parameters for α-D-glucose followed a similar approach. Since glucose is a solid at room temperature, a model of the crystal structure[36] rather than a liquid was used in addition to the gas phase, and the sublimation enthalpy instead of the vaporization enthalpy. It remains challenging to reproduce the lattice parameters for α-D-glucose. Minor adjustments in atomic charges and Lennard-Jones parameters were made to reproduce the density and the enthalpy of sublimation within the range of experimental error (Table S2). Hereby, atomic charges were assigned by analogy to chemically similar compounds (alcohols, ethers) and using boundary conditions such as local charge neutrality of functional groups (e.g., for the force field types ho + oh + c1 + hc in Figure S11, the net charge is zero). Improvements to IFF quality will require a review of the charge distribution, including stereoelectronic effects, and the representation in the force field along with a review and rationale for all other parameters.

**S1.8. Simulation of Epoxy-Amine Polymerization Using REACTER.** *Model Building, Creation of Template and Map Files.* A structure with bis[4-(3-aminophenoxy)phenyl] sulfone (*m*-BAPS), bisphenol A diglycidyl ether (DGEBA), the template for the product of the two monomers to create a secondary amine, and the template for the product of the secondary amine and DGEBA to create a tertiary amine were created in a singular Materials Studio file for use with REACTER.[2] Usage of REACTER requires that all molecules and atom types exist in the same file (i.e., product atom types that differ from reactants). Separately, individual pre-reaction molecular topologies and post-reaction topologies were created to work with the AutoMapper tool.[45] All structures were exported in .car/.mdf format and converted to a LAMMPS data file using msi2lmp.exe.

The REACTER map file was created by visually inspecting the pre- and post-reaction topology files in OVITO[46] and manually filling out the required information for map files in a text document. As such, two map files were created for the primary and secondary epoxy-amine reactions. The map files contain the “edgeIDs”, “InitiatiorIDs”, and atom ID “equivalences” between the pre- and post- reaction molecule files.

*Equilibration Phase.* The equilibration phase of the simulation assumed usual settings for molecular dynamics simulations in the NPT ensemble as described earlier, including some important differences. Some additional memory, beyond that for non-reactive MD simulations, needs to be available to accommodate special bonds, bonds, and angles that may form during the reactions. The "bond_style" used harmonic bonds and was set to "class2" since we employed IFF-R in PCFF format for this example (Morse bonds could also be used). The "improper_style" was disabled and set to "none", which avoided errors while using REACTER and the "fix bond/break" command. The newest release of LAMMPS (at the time of writing) has meanwhile added support for using "fix bond/break" with class 2 impropers.

After these initial settings, we created coordinate files for individual monomers and replicated the monomers such that the simulation cell contained more than 30,000 atoms (IFF-R can handle $>10^6$ atoms if needed). Next, the template files for pre- and post- reaction molecules were read and assigned specific names. The initial structure was subjected to energy minimization to relax the initial, randomly distributed, and loosely packed monomers. To create a bulk model system of equilibrium density, electrostatic calculations were turned off, the molecular model was subjected to MD simulation at 300 K and 1000 atm pressure for 50 ps to compress the reactants into a bulk structure. The electrostatic interactions were turned back on and the simulation cell was allowed to equilibrate in MD simulation at 300 K and 1 atm pressure for 50 ps. Turning off the electrostatic interactions during the compression into a bulk structure was necessary in LAMMPS, as otherwise the code may lose track of atoms (it is possible this is a glitch in LAMMPS since turning off electrostatic interactions in the interim is not necessary in other MD programs such as Discover, NAMD, or GROMACS). .

*Reaction Phase.* After equilibration, the barostat was unfixed and the REACTER settings were applied by using “fix bond/react”. We used the “stabilization” option with “stabilize_steps” set to 500 fs. The possibility for reactions was checked every 100 fs with the minimum distance set to 0 Å, the maximum distance set to 6 Å, and the probability set to 0.1. Additionally, the Nose-Hoover thermostat was applied to all atoms not participating in the reaction with a temperature of 600 K. A separate velocity rescaling thermostat was applied to all reacting atoms with temperature rescaling set to 200 K. The choice in thermostats ensures that non-reacting atoms can efficiently move into reactive configurations during the limited simulations times of 150 ps, and reacting atoms mitigate strong attractive forces upon bond connection as well as strong repulsive forces when related nonbonded atoms happen to have close contacts or overlap during the reaction. The reaction simulation was run for a total time of 49.95 ps. Hereby, the runtime was deliberately not divisible by the time interval REACTER checks for reactions (0.1 ps) to avoid a situation in which the simulation could conclude with unresolved reactions and errors before the next step. Overall, the REACTER method is a heuristic approach to model chemical reactions, and requires further development towards ergodicity.

*Strain and Failure Phase.* Bond parameters were reassigned to Morse bonds after the reaction phase of the simulation to ensure a continuous simulation as bonds form and break. The “bond_style” was redefined to “morse” or “hybrid morse class2” before starting the “strain phase”. Thus, Morse bond parameters should be made available at the outset. The “fix bond/react” settings were unfixed in reverse order in which they are applied, which is necessary to avoid an error. Next, the “fix bond/break” options should be applied or read in from auto_morse_bond_update.py, which will apply the disconnection of Morse bonds past a pre-defined bond elongation, e.g., 200% (Section S2). IFF-R simulations proceeded as normal MD simulations without additional options, however, multiple Nose-Hoover thermostats can also be applied as in the “fix bond/react” settings. The simulations of stress-strain curves up to failure were then applied as described in section S1.2.

## S2. A Graphical User Interface and Python Code to Automatically Convert Harmonic Bonds into Morse Bonds

A graphical user interface (GUI) and the auto_morse_bond_update.py code in the Supplementary Software (“CODE_AND_GUI_FOR_AUTO_MORSE_BOND_ASSIGNMENT”) automate the process of switching harmonic bonds to Morse bonds (Figure S1). The code provides a LAMMPS script that can be read in with the “include” statement so that “fix bond/break” commands can be quickly and easily used for each newly assigned Morse bond. The input to the code is a LAMMPS data file that has classical bond parameters. The outputs include a data file with Morse bond parameters, a LAMMPS input script file containing “fix bond/break” commands for the newly assigned Morse bonds, a pdf file showing Morse bond parameterization graphically, and a logfile with system information that can be useful for tracking parameter assignment.

To convert harmonic bonds to Morse bonds using auto_morse_bond_update.py (Figure S1), we use a graph theory-based approach with chemical intuition to find the bond order of each bondTypeID in the LAMMPS datafile. We then assign an experimental bond dissociation energy to that bondTypeID using the values found in our expandable database “src/auto_morse_bond/Morse_parameters.txt”, included in the Supplementary Software. Subsequently, the alpha parameter is fitted by minimizing the sum of residuals of the harmonic bond and the Morse bond near the equilibrium position, enabling an exact match of the Morse bond to the harmonic potential at small bond elongation. Lastly, the code writes a new LAMMPS datafile with the new Morse bond parameters in a hybrid style. Hereby, certain bondTypeIDs may remain of the harmonic type since only weak bond types that are exposed to load and prone to disconnect need to be converted to a Morse bond. Figure S1 shows an overview of the GUI and further options.

## S3. Molecular Response Upon Bond Dissociation and Recommended Bond Cutoffs

**S3.1. Molecular Strain and Topology Upon Dissociation of Morse Bonds.** When using a Morse bond potential to simulate bond dissociation, the bond, angle, torsion, improper, cross terms (if present), and nonbonded interactions in the energy expression contribute to the total energy regardless of how much the bond has been elongated apart from equilibrium. Using a Morse bond alone, also does not disconnect the originally bonded atoms nor turn off the other energy contributions. The first problem can be solved using a maximum Morse bond length ($R_{max,\ ij}$), or optionally including a probability of bond cleavage at $R_{max,\ ij}$ for each 'reactive' bond. For example, the command "fix bond/break" in the simulation code LAMMPS can be employed to disconnect exposed bonds, turn off all other associated energy contributions (angle, dihedral, etc), and enable new nonbonded interactions between the dissociated atoms, e.g., van der Waals and Coulomb interactions between formerly 1, 2 connected and 1,3 connected atoms.

In addition, the disconnection of bonds at strains relatively far from equilibrium typically releases a large amount of strain energy in the form of increased particle velocities. The sudden release of such strain energy, which can include many neighbor atoms, can be mitigated using velocity rescaling (e.g. a temperature window of 10 K), which we have used here, and which appears to be sufficient for a smooth progression of the simulation. Alternative methods to remove excess energy from the system include other thermostats, such as velocity rescaling through canonical sampling,[47] smoothly adjusting bonded interactions to zero using additional parameters, such as in the development of the AIREBO potential[48] and as utilized for the dissociation of polymers,[49] or by using REACTER with the "stabilization" option instead of "fix bond/break". However, using REACTER to break bonds will involve the creation of bond break templates which require more user effort than "fix bond/break".

**S3.2. Dissociation of Morse Bonds and Bond Cutoffs.** The dissociation of Morse bonds, for example, by using "fix bond/break" in LAMMPS, is not strictly necessary to simulate stress-strain properties of materials up to failure by bond dissociation. For a more realistic analysis at and post-failure for a chemical system, however, we propose a shifted Morse potential with a bond cutoff at about 200% of initial bond length. Systematic testing of various cutoff distances for polymers and carbon nanotubes suggests this value, which also is close to the inflection point of the Morse bond force vs bond length curve. The inflection point of a Morse bond force vs bond length curve represents the point at which the attractive force of a Morse bond is overcome by the strain energy. The bond energy at this cutoff distance then reaches zero, allowing a relatively unbiased analysis of energy changes in 3D atomic structures upon failure, especially when the strain continues far beyond the strain at break (Figure 1 and Figure S2).

The default Morse potential (left side of Figure S2a) is not suited as it introduces an energy discontinuity upon bond scission (e.g., using the "fix bond/break" command), as the bond energy then drops from $D_{ij}$ to 0. A shifted Morse potential (right side of Figure S2a), on the contrary, minimizes discontinuities in bond energy. The implementation of the shifted Morse potential is possible, with moderate effort, in any MD program. As an example, LAMMPS requires modification of the source code (provided in the Supplementary Software as "user Morse").

Hereby, we found that the shifted Morse potential may not need to result exactly in zero energy at the cutoff distance and may be applied for a range of bond cutoff distances $R_{max,\ ij}$. For several tested examples, we obtained similar results, however, with some dependence on the bond cutoff distance (Figure 2b-e). Once $R_{max,\ ij}$ is sufficiently large, the results converge (Figure S2b-e and Figures S3 to S5). The minimum bond cutoff distance depends on the type of bond, in our instances of CNTs and polyacrylonitrile, it was between 140% and 170% of the equilibrium bond length. On the other hand, bond cutoff distance is unlikely larger than 2 or 3 times the original bond distance as then the remaining localized electron density for covalent bonding goes to zero. Other remaining bonded energy terms such as angles, dihedrals, and impropers increasingly lose influence and become unphysical, and nonbonded interactions between the dissociating atoms and connected groups of atoms begin to play a role. Therefore, we recommend 200% of the original bond length as a cutoff distance, and no cutoffs larger than 300% of the original bond length.

In more detail, arbitrarily selecting a "fix bond/break" cutoff distance can cause premature failure and misleading mechanical properties (Figure S3). For example, using a single polyacrylonitrile (PAN) chain with a 1.54 Å equilibrium C-C bond distance ($r_0$), we found that failure occurred at lower strain values for cutoffs less than 2.6 Å and remained unchanged for cutoff distances of 2.6 Å and higher. The total energy of the PAN tensile simulation spikes prematurely for cutoffs less than 2.6 Å and shows a large irregularity for even shorter bond cutoffs such as $R_{max,\ C\text{-}C}$ = 1.7 Å (Figure S3a, c). Small spikes in the total energy at the strain of bond breaking are commonly observed for any bond cutoff, whereby the total energy as a function of strain remains unchanged for cutoffs greater than 2.6 Å, equal to >170% of the original bond length. The spikes are likely the result of strain energy release from an overall strained backbone, and the simultaneous addition of nonbonded interactions between the formerly bonded atoms.

A SWCNT was used as another example for comparing the effect of bond cutoff distances on the stress-strain curves and energies (Figures S4 and S5). Early onset of failure is shown for $R_{max, C\text{-}C}$ values less than 2.0 Å (Figure S4a) and, like PAN, premature spikes in the total energy were observed for even shorter bond cutoff distances (Figure S3c). Values of $R_{max}$ greater than or equal to 2.0 Å show nearly identical, overlapping stress vs strain curves as well as curves of total energy vs strain (Figure S3b, d). There appears to be no upper limit for a bond cutoff up to 4.0 Å, and we recommend $R_{max}$ to be set at ~200% of the original bond length $r_0$. If a shorter cutoff is desired, it should be at least 170% of the length of a C-C single bond (2.6 Å) for polymers such as PAN and at least 140% of the length of an aromatic C-C bond such as graphite.

The magnitude of the spikes in total energy at bond dissociation is susceptible to the shift of the Morse potential and the new nonbonded parameters for the reaction products (atomic charges and LJ parameters), and not significantly to the cutoff distance $R_{max}$ in the recommended range. A reduction of the small energy spikes at bond breaking, if desired, is thus possible by modification of the shift of the Morse potential, which can be chosen different from $D_{ij}$. Smoothing functions with additional empirical parameters could also be employed.

**S3.3. Necessity for a Shifted Morse Potential and Analysis of Individual Energy Contributions During Failure.** The spontaneous spikes in energy upon failure, as shown for PAN (Figure S3), and SWCNTs (Figure S4) were further analyzed in terms of the individual energy contributions for the example of SWCNTs (Figure S5). Using the "fix bond/break" command (*Rmax* = 2.84 Å for CNT) with or without shifting the Morse potential, allows continued simulations of dissociated SWCNTs beyond the failure strain of 20% to higher strains (e.g., >40%). When not allowing for bond breaking (with or without a shifted Morse potential), the simulations become unstable and terminate once the bonds are separated by multiples of the equilibrium bond lengths. Termination can result from atoms that participate in the bonds/angles/dihedrals involved at the site of dissociation and migration outside the range used in spatial decomposition with multi-processor computing (becoming “lost”).

Using the unshifted Morse potential, the total potential energy, bond energy, and angle energy return to a baseline value after bond dissociation after reaching the bond cutoff distance (Figure S5a-d). A spike in kinetic energy is observed upon failure at 20% strain in all instances, regardless of whether the bonds are subsequently disconnected or not, due to a gain in atom mobility (Figure S5b). Bond and angle energies are lower when permitting bond disconnection compared to not permitting bond disconnection due to the opportunity for relaxation (Figure S5c, d). Using the modified (shifted) Morse potential, the potential energy and bond energy are higher than the baseline value after dissociation without a shift of the Morse potential. The difference equals the product of the number of bonds dissociated and the bond’s dissociation energy.

The van-der-Waals energy shows a spike and then decreases (Figure S5e). Van-der-Waals interactions for severed bonds are turned on upon bond disconnection and become attractive, as is seen by the decreasing energy when compared to the examples where “fix bond/break” was not used (Figure S5e). The electrostatic energy increases (Figure S5f) when permitting bond disconnection due to switching on previously excluded electrostatic interactions (and retaining atomic charges after bond breaking for simplicity here). The higher electrostatic energy after bond dissociation for CNTs, as an example, results from the negative charge on the virtual π-electrons and the positive charges on the carbon atoms, which repel each other after failure. Other contributions from improper energy and cross terms (not shown in Figure S5) also increase and lead to an overall higher plateau of the total potential energy after bond dissociation when using the modified Morse bond potential with the "fix bond/break" command than without "fix bond/break" (Figure S5a).

**S3.4. Chemical Reactions After Bond Dissociation.** From a chemistry perspective, bond scission results in the formation of radical or charged species, followed by secondary reactions depending on the composition and state variables of the system (Figure S6). Existing knowledge of reaction mechanisms and theory need to be used to guide the simulation towards realistic reaction outcomes. The newly formed chemical species need to be represented following the equation for the chemical reaction, associated stoichiometries, and by updated force field parameters. For example, new atom types and associated parameters may be assigned to the reaction products, which may often be available, or intermediate reactive chemical species can be represented by dedicated new atom types and force field parameters (radicals, radical ions, carbocations, carbanions, etc.). The latter scenario can be more challenging; however, it can be achieved using IFF protocol and high-level quantum mechanics.

For the test cases presented here, atom types and force field parameters other than charge remained the same after dissociation. It is unlikely that dissociated atoms spontaneously form bonds on-the-fly during a tensile simulation at high strain rates typical of molecular dynamics simulations. For more detailed possible reactions, some pathways are suggested in Figure S6 and can be implemented as heuristic changes, e.g., as implemented in REACTER. Upon new charge assignments, overall charge neutrality of reaction sites needs to be maintained. Otherwise, automatic protocols may unintentionally create systems with a net charge and unphysical results.

For CNTs and other graphitic materials, bond dissociation under tension is likely homolytic and two radicals will form. In this scenario, assigning new charges (zero) and adjustments to the Lennard-Jones well depth (epsilon) would seem reasonable. Follow-on reactions with $H_2O$, $O_2$, $N_2$, or other chemical species present in air may need to be defined. Heterolytic cleavage of C-O bonds is a likely scenario for bond scission in cellulose. Two new force field types would be suitable for the positively charged carbon and negatively charged oxygen atoms. Intermediate parameterizations of this kind could be avoided if the products are assumed to immediately stabilize as a $sp^2$ carbon or hydroxyl group which already have force field parameters defined in PCFF-R or other suitable force fields (Figure S6). Similarly, the failure of PAN would require product specifications or new nonbonded parameters for radicals or carbocations.

In summary, chemical reactions are diverse and follow-on reactions need to be considered case-by-case taking into account existing knowledge and available experimental data.

**S3.5. Custom LAMMPS Option for a Shifted Morse Bond Potential and IFF-R.** For easy use of IFF-R, we developed a custom LAMMPS option for IFF-R that can be installed when compiling LAMMPS ("CODE_FOR_CUSTOM_LAMMPS_OPTION_USER_MORSE-2Aug23" in the Supplementary Software). When installing custom options during compilation, the command

"make yes-user-morse"

should be issued. The new LAMMPS option defines a Morse potential with a shift such that the bond energy can reach 0 kcal/mol at the "fix bond/break" cutoff distance, which was not previously available and is needed for monitoring bond energy in bond breaking simulations in high accuracy and compatibility with IFF-R, IFF, CHARMM, AMBER, CVFF, PCFF, COMPASS, and other force fields for multiple materials classes as described. The syntax for the IFF-R Morse bond potential is as follows:

```
bond_style    morse
```

bond_coeff i d0 alpha r0 [rcut [offset]]

The potential form of this equation is as follows:

$$E_{Morse} = d0 * \left(1 - \exp\left(-alpha * (r - r0)\right)\right)^2 - E_{rcut}[-offset] \tag{S3}$$

In the command, $i$ specifies the index of the bond type in the LAMMPS data file, d0 the bond dissociation energy in kcal/mol, alpha the width of the potential, and r0 the equilibrium distance in Å (Figure 1). $E_{rcut}$ specifies the bond energy at the bond cutoff distance *rcut*, e.g., at 200% of the original bond length, and shifts the Morse bond potential to zero at *rcut* automatically. However, *offset* enables a manual shift of the Morse potential that allows further adjustment of the bond energy towards the value of *d0* (*d0* > $E_{rcut}$). For typical use, no offset is needed (*offset* = 0).

When *rcut* is not specified, the bond energy is not offset and the potential energy behaves similar to the original Morse bond energy potential. When *rcut* is specified but not the *offset*, the Morse potential is shifted by the energy of the bond at the bond cutoff distance ($E_{rcut}$) which is internally determined, alleviating the user from the burden of calculating the cutoff energy and the needed offset. Otherwise, if the energy *offset* is specified by the user, the specified offset is used instead. In this manner, the user can freely customize the shift in the Morse potential energy. For example, by shifting it manually so that a bond energy of 0 kcal/mol is reached at the "fix bond/break" *rcut* distance, to the bond dissociation energy (*d0*) by specifying *d0* as the *offset*, or by anything else the user desires with proper justification.

**S4. Bond Order Analysis of Aromatic Amines Using Resonance Structures**

The high-level quantum mechanical bond scans (MP2) show an interesting correlation between the bond order and the dissociation energy (Figure 2). The relative dissociation energies for the same type of bond with different substituent patterns can be qualitatively described through the resonance structures, shown here for a DETDA molecule and a 4, 4'-DDS molecule (Figure S7). For DETDA, we counted the sum of all single bond versus all double bond representations for identical bonds including molecular symmetry (1, 2 and 2, 3; 3, 4 and 1, 6; 4, 5 and 5, 6) (Figure S7a). The statistics indicates that the double bond character increases in the order 3,4 < 1, 2 < 5, 6, consistent with the results by MP2 calculations, especially up to 150% of the original bond length (Figure 2a, b).

A similar analysis for 4, 4'-DDS reveals a push-pull effect of electron density between the amino groups and the sulfonate groups in para position (Figure S7b). The 2,3 carbon-carbon bonds in 4,4'-DDS have almost full double bond character and are ~30 kcal/mol stronger than the 1,2 and 3,4 bonds (Figure 2c, d).[50] Stereoelectronic effects, such as double bond character and related differences in bond dissociation energies, can be incorporated into IFF-R and other force fields (PCFF-R, CHARMM-R, AMBER-R, OPLS-R) using the Morse bond parameters.

**S5. Analysis of Mechanical Properties of Defective CNT Structures and of Spider Silk Protein (Spidroin)**

Understanding the role of defects and of order versus disorder in materials across scales on their functions is a continuing challenge in materials science. IFF-R can contribute to understanding non-linear behavior under deformation and failure. We illustrate some advanced calculations of the role of defects in CNTs (Figure S8) and of the failure of the spider silk protein spidroin (Figure S9). 4 models of a SWCNT were created with an increasing number of vacancy defects (Figure S8a). Calculations of the stress-strain curves indicate that a single vacancy reduces the tensile strength by approximately 15% (Figure S8b). Subsequent vacancies lead to a less pronounced impact on strength. At the same time, the modulus at low strain and the initial shape of the stress-strain curve are not noticeably affected by the presence of one or multiple vacancies. Larger defects, such as Stone-Wales, broken ends in CNT bundles, and defective multilayer graphitic structures in carbon fiber can also be investigated, which is beyond the scope of this work.

At times, it is desirable to predict the temperature-dependent mechanical behavior, especially for materials subjected to extreme environments. We considered the elasticity and failure mechanism of spider silk protein (spidroin), which contains 155 amino acids per strand and is approximately 77.5 nm long (assuming each amino acid is roughly 0.5 nm in width) at two different temperatures (Figure S9). This example shows how IFF-R can be applied to biomolecular force fields such as CMARMM36, AMBER/GAFF, and biochemical problems (Figure S9). We built a model containing 2 spidroin protein strands using CHARMM36m (Figure S9a) and subjected the models to stress-strain simulations at room temperature (298 K) and at a temperature of 1 K near absolute zero for comparison (Figure S9b, c). At 298 K, entanglements of the long protein strands occurred and led to bond breaking after high elasticity and ductility (~400% engineering strain). The elastic modulus, near zero strain, was computed to be ~3.6 GPa. These simulated properties agree with experimental observations[4-7] and could be explored further for larger model sizes, multiple configurations, spidroin mutations and different amino acid sequences (Figure S9b). In contrast, at a temperature of only 1 K, the mobility of the atoms and of the protein strands was two orders of magnitude more restricted. Based on our start structure, entanglements did not form, and the strands slid apart without bond failure upon applied stress (Figure S9c). The tensile strength, or shear strength, was then lower at approximately 0.7 GPa (Figure S9b, c), and the elastic modulus was somewhat higher at 1 K than at 298 K.

Spidroin would likely demonstrate brittle fracture at temperatures near absolute zero, e.g., if using a model of a supercooled, entangled spidroin protein (e.g. the start structure at 298 K after equilibration). If placed under strain at 1 K, stress concentration at the entanglements would lead to failure since few bonds are initially aligned in the strain direction and the relaxation time towards alignment in response to strain would be very long at 1 K versus quite short at 298 K. However, we used aligned strands and the timescale of the simulation allowed lateral motion with an even stress concentration, also showing a dependence of the failure mechanism on the initial structure (Figure S9c).

**S6. Comparison of Computational Speed for Simulations of Large Systems**

The speed up by IFF-R for large systems of 100,000 to 200,000 atoms relative to ReaxFF is about the same as for smaller systems with 1,000 to 5,000 atoms, i.e., 20-30x for polymers, 50x for metals, and 8x for CNT bundles (Figure S10 and Figure 4). In addition, simulations with ReaxFF are at least 10x more memory-intensive than simulations with IFF-R and could not be started with less than 12 CPUs due to insufficient memory (4.84 GB/CPU available). In comparison, all large systems could be run on 1 CPU using IFF-R, which is an added benefit for resource efficiency. For a comparison on equal footing, all large systems were run using 24 CPUs for both IFF and ReaxFF.

The IFF-R simulations would finish within 24 to 48 hours on the 24 CPUs, whereas the ReaxFF simulations could take up to a month. For many supercomputer job schedulers running a single computation for 24-48 hours is allowable while running a single job for a month is typically prohibited and not a preferred timeframe for many studies. Therefore, IFF-R can also serve resource-limited researchers to simulate mechanical properties and failure of complex materials. Foremost, IFF-R achieves highest accuracy and interpretability among alternative MD models.

### S7. Surface Energy and Vaporization Enthalpy of Various Materials Classes with IFF-R Versus Available ReaxFF Parameter Sets

The computed surface energies and vaporization energies, respectively, for graphite, γ-iron, propionitrile, and α-D-glucose using IFF-R and ReaxFF were compared relative to laboratory measurements (Table 3). For graphite, virtual π-electrons in IFF-R are essential to reproduce surface, wetting, and adsorption properties simultaneously due to associated multipole moments.[51,52] The computed surface energies or cleavage energies, respectively, agree with well-known experimental reference data within the experimental uncertainty of ~1% (Table 3).[18,19,53] Simulations of graphite with ReaxFF, which do not include virtual π-electrons, lead to +30% deviation in cleavage energy utilizing the Liu et al. (ref. [17]) parameter set.

For the (111) γ-iron surface, both the IFF-R and ReaxFF parameters reproduce experimental values of the surface energy well with 1% and 6% deviation, respectively.[54-56] However, the ReaxFF parameters used in this simulation (Shin et al.[57]) are different from those used in the stress-strain simulation (Aryanpour et al.[58]) to achieve a better fit (Figure 3h and Table 1). The Aryanpour ReaxFF parameters (ref. [58]) would result in greater than 200% deviation of the computed surface energy from experimental reference data when used. The differences demonstrate that using the same ReaxFF parameter set for iron, deviations on the order of 100% in either surface properties or mechanical properties are inevitable, which is not the case with IFF and IFF-R.

For proprionitrile, the IFF-R parameters reproduce experimental data for the vaporization energy, and the modified PCFF-R parameters replicate the enthalpy of sublimation of glucose (Table 3)[20,59,60] while needing major improvements to reproduce the crystal structure (Section S1.7). ReaxFF parameters perform well for glucose but not for propionitrile, which shows 30% deviation in vaporization energy (Table 3). We tested both the Liu[17] (Table 3) and Damirchi[16] ReaxFF parameter sets (Table S2). The Liu parameters provide better nonbonded properties and can be used in calculations with the *l*g van-der-Waals correction. The *l*g van-der-Waals correction is an additional vdW interaction more adequate to capture long-range dispersion forces and to reproduce mass densities in molecular crystals than the standard Morse potential in typical ReaxFF potentials. The Liu parameters were also optimized using the enthalpy of sublimation of solids and appear to be overall a better fit to compare surface energy and vaporization enthalpies to IFF-R results. Damirchi parameters for ReaxFF lead to higher combined deviations of 11% for propionitrile and 32% for glucose (Table S2).

In more detail, ReaxFF exhibits deviations in the surface energy of graphite and in the vaporization energy of organic liquids of up to 30% using two different parameter sets by Liu and Damirchi (Table S2). The errors are similar overall yet have a different distribution. The parameter set by Liu[17] includes long range van-der-Waals corrections, which may be considered an improvement over the parameter set by Damirchi,[16] although not showing lower errors. Improvements of ReaxFF to reduce uncertainties of 10% to 30% to IFF and IFF-R standards of ~5% would require a time-consuming reparameterization or may not be possible without enlarging errors for other chemistries covered by the same parameters.

The Damirchi parameters of ReaxFF[16] overpredict the surface energy of graphite with about 11% deviation from experimental data, which is an improvement over the Liu parameters[17] with 30% deviation. The improvement originates from optimization for flattened carbon nanotubes, informed by IFF-PCFF,[16,61] that are similar in structure and composition to graphitic layers. The Damirchi parameters also predict the enthalpy of vaporization of propionitrile with about 11% deviation from the experimentally determined value, which is better than 36% deviation with the vdW-corrected Liu parameters. Conversely, the Liu parameters for ReaxFF predict the enthalpy of sublimation of α-D-glucose with only 4% deviation, which is much better than the Damirchi parameters with more than 30% deviation from experiment. The choice of ReaxFF parameter set is therefore often a trade-off that involves significant errors somewhere. In comparison, deviations with IFF-R are smaller and remain below a maximum of 3% (Table 3).

**Supplementary References**